\documentclass[draftclsnofoot, onecolumn, comsoc, 12pt]{IEEEtran} 
\IEEEoverridecommandlockouts
\usepackage{graphicx}
\usepackage[group-separator={,},group-minimum-digits=4]{siunitx}
\usepackage{lipsum}
\usepackage{algorithm}
\usepackage{color, colortbl}
\definecolor{Gray}{gray}{0.9}
\definecolor{White}{gray}{1}
\usepackage{makecell} 
\usepackage{array}
\newcolumntype{P}[1]{>{\centering\arraybackslash}m{#1}}
\newcolumntype{A}{>{\columncolor{White}}P}
\usepackage{algpseudocode}

\usepackage{cite}
\usepackage{subfigure}
\usepackage{multirow} 
\usepackage{amsmath,amssymb,amsfonts}
\usepackage{float}
\usepackage{amsthm}
\usepackage{graphicx}
\usepackage{textcomp}
\usepackage[utf8]{inputenc}
\usepackage[english]{babel}
\usepackage{booktabs}
\usepackage[noabbrev]{cleveref}
\usepackage { optidef }
\def\BibTeX{{\rm B\kern-.05em{\sc i\kern-.025em b}\kern-.08em
    T\kern-.1667em\lower.7ex\hbox{E}\kern-.125emX}}

\newtheorem{remark}{Remark}

\def\tcb{\textcolor{black}}

\begin{document}
\bstctlcite{IEEEexample:BSTcontrol}
\title{A Bistatic ISAC Framework for LEO\\Satellite Systems: A Rate-Splitting Approach}

\author{Juha Park, Jaehyup Seong, Jaehak Ryu, Yijie Mao, and Wonjae Shin \vspace{-5mm}
    \thanks{
    }
    \thanks{This
article was presented in part at the IEEE International Conference on Communications Workshop (ICC Workshop), Denver, USA, June 2024 \cite{MyICCPaper}.\\J. Park, J. Seong, J. Ryu, and W. Shin are with the School of Electrical Engineering, Korea University, Seoul 02841, South Korea 
    (email: {\texttt{\{juha, jaehyup, jaehak99, wjshin\}@korea.ac.kr}});
    Y. Mao is with the School of Information Science and Technology, ShanghaiTech University, Shanghai 201210, China (email: {\texttt{maoyj@shanghaitech.edu.cn}}).
    }} 
\maketitle

\begin{abstract}
Aiming to achieve ubiquitous global connectivity and target detection on the same platform with improved spectral/energy efficiency and reduced onboard hardware cost, low Earth orbit (LEO) satellite systems capable of simultaneously performing communications and radar have attracted significant attention. Designing such a joint system should address not only the challenges of integrating two functions but also the unique propagation characteristics of the satellites. To overcome severe echo signal path loss due to the high altitude of the satellite, we put forth a bistatic integrated sensing and communication (ISAC) framework with a radar receiver separated from the satellite. For robust and effective interference management, we employ rate-splitting multiple access (RSMA), which \tcb{splits and encodes} users' messages into private and common streams. We optimize the dual-functional precoders to maximize the minimum rate among all users while satisfying the Cramér-Rao bound (CRB) constraints. Given the challenge of acquiring instantaneous channel state information (iCSI) for LEO satellites, we exploit the geometrical and statistical characteristics of the satellite channel. To develop an efficient optimization algorithm, semidefinite relaxation (SDR), sequential rank-1 constraint relaxation (SROCR), and successive convex approximation (SCA) are utilized. Numerical results show that the proposed framework efficiently performs both communication and radar, demonstrating superior interference control capabilities. Furthermore, it is validated that the common stream plays three vital roles: i) beamforming towards the radar target, ii) interference management between communications and radar, and iii) interference management among communication users. 
By leveraging the multi-functionality of the common \tcb{stream} and optimizing resource allocation, our approach \tcb{enables} efficient joint operation of satellite sensing and communication even without a dedicated radar sequence.

\end{abstract}


\section{Introduction}
Low Earth orbit (LEO) satellite communications have recently received considerable attention from academia and industry \cite{kodheli2020satellite}. The advantages of LEO satellites lie in their relatively low altitude ($\num{300}$-$\num{2000}$ km) compared to geostationary Earth orbit (GEO) and medium Earth orbit (MEO) satellites, resulting in reduced latency and higher data rates. Furthermore, recent technological advancements make the LEO satellite industry more cost-effective and make it easier to  
launch satellite constellations that provide global coverage, even in areas where traditional terrestrial networks struggle to provide service. Meanwhile, the proliferation of devices that utilize electromagnetic waves is exacerbating spectrum congestion. As a result, communications and radar technologies are competing for the S-band (2-4 GHz), C-band (4-8 GHz), and possibly millimeter wave (mmWave) bands \cite{xu2021rate}. Motivated by the aforementioned challenge, integrated sensing and communication (ISAC), also referred to as dual-function radar communications (DFRC), joint communication and radar (JCR), or radar-communications (RadCom), has emerged as a promising solution to alleviate spectrum congestion and lower hardware costs \cite{liu2022integrated}. As LEO satellites are expected to play a key role in future communications systems, such as beyond 5G (B5G) and 6G, they are not immune to this problem. Besides, the relatively low altitude of LEO satellites offers a great potential for ubiquitous sensing capabilities. In this context, it is worth exploring ISAC for LEO satellite systems (LEO-ISAC) capable of simultaneously providing global connectivity and sensing capabilities. Although a plethora of research has investigated terrestrial ISAC systems \cite{liu2020joint, terrestrial_ISAC_4, gao2023cooperative, chen2022generalized, pucci2022performance, kanhere2021target}, their direct application to LEO satellite systems is not straightforward due to the different signal propagation characteristics \cite{you2022beam, liu2024max,yin2022rate}. Specifically, the high-speed mobility and the long propagation distance of the LEO satellites pose the challenge of severe radar echo path loss (typically proportional to the square of the distance) and the difficulty of obtaining accurate and timely channel state information (CSI) at the satellites.

In this paper, we focus on designing a suitable system capable of addressing the challenges of implementing ISAC for LEO satellite systems, with particular emphasis on the signal processing perspective.


\begin{figure*}[!ht]
\centering
 		\includegraphics[width=0.9\linewidth]{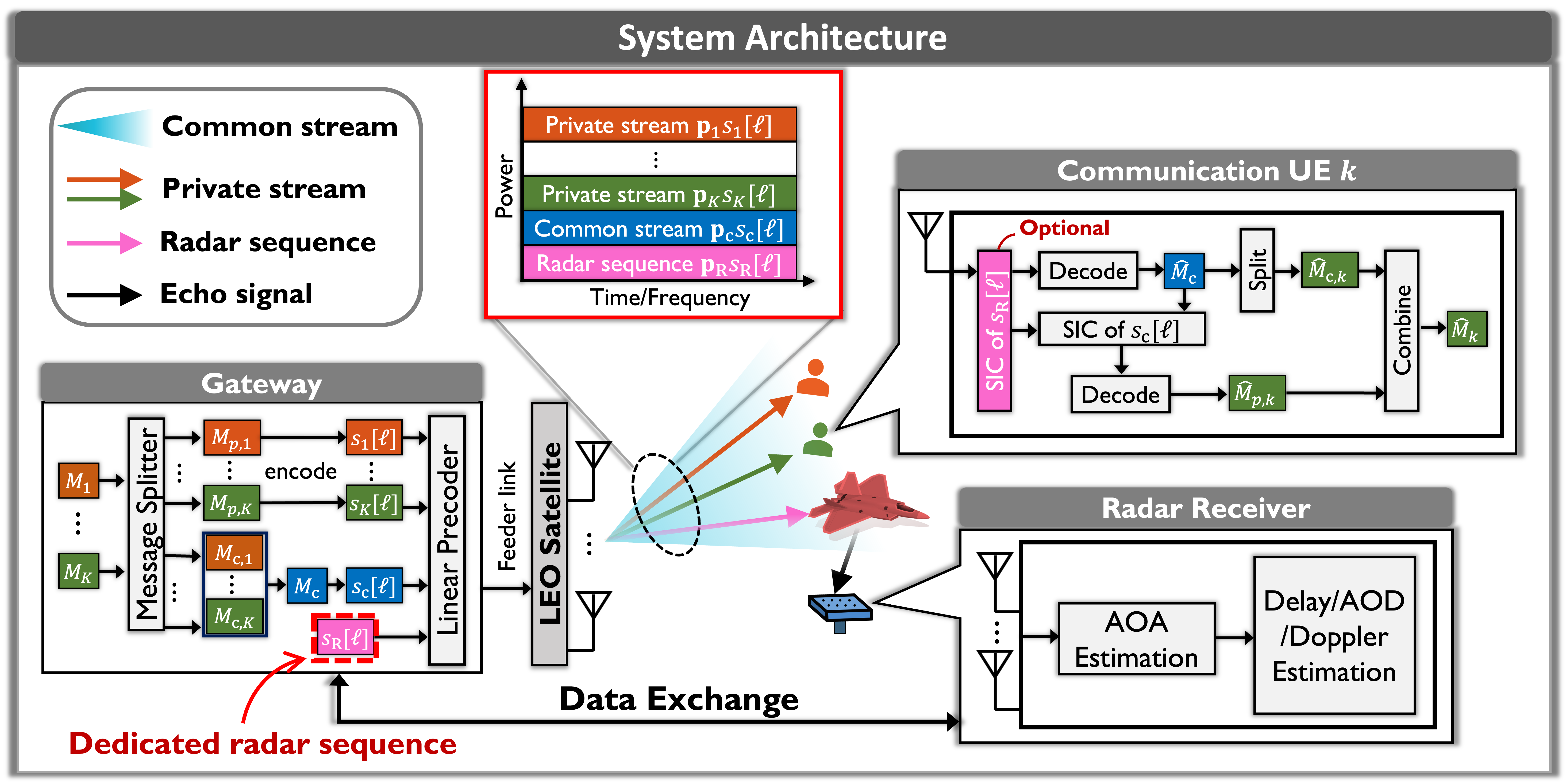}
 		\caption{Overall system architecture of the proposed bistatic LEO-ISAC framework.}
    	\label{block_diagram}\vspace{0mm}
\end{figure*}
\subsection{Literature Review}
Full frequency reuse is a promising strategy to improve communication performance by efficiently utilizing limited spectrum resources for LEO satellite networks \cite{kodheli2020satellite, lee2023coordinated, you2020massive}. This scheme uses a single frequency band instead of dividing the available frequency band into multiple orthogonal sub-bands. However, this approach leads to severe co-channel interference among communication users. To address this issue, researchers have explored precoder design techniques relying on multiple-input multiple-output (MIMO) technology \cite{you2020massive, lee2023coordinated, yin2020rate}. In parallel, ISAC, where the communication and radar functions share the same time-frequency resources using MIMO technology, has been extensively studied. {By doing so, the LEO satellite communications and the ISAC have been evolving with the common goal of improving radio resource efficiency.} This naturally leads to a research direction combining LEO satellite communications and ISAC technologies to achieve further advanced resource efficiency. An overview of the wide range of related work, ranging from terrestrial ISAC to satellite-based ISAC, is presented as follows:

\textbf{\,Terrestrial ISAC:}
Much of the ISAC research to date has been devoted primarily to terrestrial systems. In \cite{liu2020joint}, the authors provided an overview of ISAC use cases and signal processing techniques. For the joint design of communication and radar functions, an appropriate dual-functional precoder is required. To achieve this, various communication performance metrics (e.g., signal-to-interference-plus-noise ratio (SINR) \cite{terrestrial_ISAC_4}, weighted sum rate \cite{xu2021rate}, energy efficiency (EE) \cite{liu2022integrated, liu2024max}, etc.) and radar performance metrics (e.g., beampattern matchness \cite{xu2021rate,liu2022integrated}, Cramér-Rao bound (CRB) \cite{terrestrial_ISAC_4, yin2022rate, gao2023cooperative, liu2024max}, sidelobe suppression \cite{9424454}, signal-to-clutter-plus-noise ratio \cite{chen2022generalized}, etc.) are jointly considered. Most ISAC research has managed interference between multiple users and radar target(s) in a spatial-division multiple access (SDMA) manner, which treats interference completely as noise. To achieve advanced interference management, \cite{xu2021rate} and \cite{gao2023cooperative} explored rate-splitting multiple access (RSMA)-based ISAC systems. In RSMA, each user's messages are split into private and common streams. Such rate-splitting concept of RSMA allows partial decoding of interference and partial treatment of interference as noise \cite{park2023rate, 9831440}, making it a powerful and robust transmission strategy to effectively control both inter- and intra-function interference for future ISAC systems.\footnote{Throughout this paper, inter-function interference refers to the interference caused to communication users by radar functions, as well as interference between uplink communication signals and radar echo signals at the satellite. On the other hand, intra-function interference refers to co-channel interference among multiple communication users.} In addition, \cite{chen2023rate} and \cite{dizdar2022energy} extended RSMA-based ISAC to multiple targets and low-resolution digital-to-analog converters (DACs) scenarios, respectively. In terms of sensing structure, ISAC research typically assumed a monostatic ISAC structure, in which a transmitter and a receiver for the radar function are physically co-located. However, \cite{pucci2022performance} and \cite{kanhere2021target} explored bistatic ISAC systems comprising a transmitter and a receiver that are separated by a distance to address issues inherent in monostatic setups, such as interference between the uplink communication signals and the radar echo signals.

\textbf{\,Satellite-based ISAC:}
Prior to research on satellite-based ISAC systems, the radar community explored the potential of using communications satellites as illuminators for passive radar systems \cite{daniel2017design, sayin2019passive, blazquez2022passive}. In \cite{daniel2017design}, the feasibility of a satellite-based passive radar system with a ground radar receiver was confirmed theoretically and experimentally. In addition, \cite{sayin2019passive} and \cite{blazquez2022passive} studied passive radar systems that utilize commercial LEO communications satellites, such as Starlink. However, these studies treated the satellites as non-cooperative sources of illumination in passive radar systems. By co-designing communication and radar functions within a single satellite, we can build potential synergies and integrate these capabilities into a \emph{dual-functional satellite}, so-called satellite-based ISAC.

Building on the synergistic gain, research on satellite-based ISAC has recently emerged, with a particular focus on LEO-ISAC \cite{you2022beam, yin2022rate, liu2024max}. The LEO-ISAC differs from satellite-based passive radar in that it involves cooperative and jointly designed communication and radar functions. The design of LEO-ISAC systems should address the following challenges: \emph{i) severe echo path loss} due to the high altitude of satellites, \emph{ii) difficulty in obtaining instantaneous CSI (iCSI)} because of the long feedback delay and the short channel coherence time, in addition to the general issues of \emph{iii) inter-/intra-function interference} in ISAC. In \cite{you2022beam}, the authors explored the LEO-ISAC and proposed an SDMA-based hybrid precoding scheme to maximize the EE while satisfying the beampattern matching constraint in the absence of iCSI. More recently, \cite{liu2024max} proposed an RSMA-based precoder design to maximize the minimum EE among all users while satisfying the CRB constraint with low-resolution DACs and perfect iCSI. Although both studies focused on the efficient use of limited power resources, they overlooked the significant power consumption associated with radar sensing tasks due to the long propagation distance of LEO satellites. In particular, both studies assumed a monostatic structure, which results in significant echo path loss primarily due to the high round-trip propagation distance, especially for relatively low-altitude radar targets such as drones, aircraft, etc. This necessitates the use of very high transmit power at the satellite, which is a critical challenge given the limited power resources of the satellite.
To tackle this challenge, a \emph{bistatic ISAC} system can be a promising solution; however, the bistatic ISAC structure induces a more complicated geometry than the monostatic ISAC structure. Specifically, the angle-of-departure (AOD) and angle-of-arrival (AOA) of the echo signal are the same in a monostatic setup, but they are quite different in a bistatic setup. In \cite{yin2022rate}, the authors simply extended existing monostatic ISAC to bistatic LEO-ISAC by assuming that the LEO satellite is always positioned at the zenith with respect to the radar receiver, which is not valid in general LEO satellite geometry.\footnote{This scenario is possible only when the satellite is positioned in GEO and the radar receiver lies on the equator at the nadir point, thus the elevation angle of the satellite with respect to the ground radar receiver is always $90^\circ$.}
This is mainly because the elevation angle with respect to the ground radar receiver should vary from $0^{\circ}$ to $90^{\circ}$ over time due to the fast movement and trajectory of the LEO satellite.
Additionally, considering a linear array radar receiver simplifies geometry but restricts the 3D sensing capabilities. Based on such an impractical system model, they proposed the design of an RSMA-based \tcb{precoders} assuming perfect iCSI. In summary, existing LEO-ISAC studies \cite{you2022beam,liu2024max,yin2022rate} failed to \emph{comprehensively} address the aforementioned three challenges in a unified framework. Specifically, \cite{you2022beam} only addressed the second challenge (lack of iCSI); \cite{liu2024max} only tackled the third challenge (inter-/intra-function interference); \cite{yin2022rate} treated the third challenge and incompletely dealt with the first challenge (severe echo path loss).
\subsection{Contributions}
In this paper, we propose a novel LEO-ISAC framework to overcome the intertwined challenges of LEO-ISAC. The key contributions of this paper are summarized as follows:
\begin{itemize}
    \item We propose bistatic LEO-ISAC systems in which the LEO satellite simultaneously provides communication services to multiple users while also performing radar functions, using the same time-frequency resources as illustrated in Fig. \ref{block_diagram}. By introducing a radar receiver separated from the satellite, the echo path loss is dramatically reduced, thereby alleviating the power consumption burden on the satellite. We employ 1-layer downlink RSMA, capitalizing on its advanced capabilities for controlling both inter-function and intra-function interference.
    
    \item We derive the CRB based on our system model and formulate the optimization problem for the RSMA-based dual-functional precoder, which aims to maximize the minimum rate among all users while satisfying the CRB constraints to ensure satisfactory radar performance. Given the challenge of obtaining iCSI on LEO satellites, we exploit the statistical and geometrical characteristics of the satellite-to-user channel. To address the non-convexity in our optimization problem, we develop an efficient two-layer (inner and outer) iterative algorithm, where the outer iteration involves semidefinite relaxation (SDR) and sequential rank-1 constraint relaxation (SROCR) and inner iteration is associated to successive convex approximation (SCA). In particular, by virtue of the SROCR algorithm, instead of simply dropping the rank-1 constraint to avoid its non-convexity, we systematically force the solutions toward rank-1 matrices in each iteration. Hence, we effectively mitigate the loss of optimality by means of such a tighter relaxation.
    \item Since our proposed bistatic LEO-ISAC differs from the conventional ISAC, an appropriate radar estimation is required. To fill this gap, we present a parameter estimation that is specifically tailored to our system model. These include a traditional multiple signal classification (MUSIC) algorithm for AOA estimation and a novel algorithm for joint delay-Doppler-AOD estimation.
    \item We numerically show a significant reduction in echo path loss with the bistatic ISAC structure compared to the monostatic one. We also validate the superior interference management capabilities of the proposed framework. Furthermore, we confirm that the common stream in our proposed LEO-ISAC plays three key roles as follows: \emph{i) forming a beam towards the radar target, ii) controlling interference between radar and communication functions,} and \emph{iii) managing interference among communication users.} The multi-functionality of the common stream enables efficient joint operations under limited resources, even without a dedicated radar sequence.
\end{itemize}

\subsection{Notations} 
Herein, standard letters, lower-case boldface letters, and upper-case boldface letters indicate scalars, vectors, and matrices, respectively. The imaginary unit is defined as $j\triangleq\sqrt{-1}$. Notations $(\cdot)^{\sf{T}}$, $(\cdot)^{\sf{H}}$, $\Vert\cdot\Vert_2$, and $\Vert \cdot \Vert_{\sf{F}}$ identify the transpose, conjugate transpose, $\ell_2$-norm, and Frobenius norm, respectively. The notation $\mathbf{I}_N$ represents an $N \times N$ identity matrix; $\mathbf{0}_{M\times N}$ is an all-zero matrix of size $M\times N$. $\mathrm{vec}(\cdot)$ denotes a column-wise vectorization operator. $\mathrm{diag}\left(a_1, a_2, \cdots, a_N\right)$ is diagonal matrix with diagonal elements $a_1, \cdots, a_N$. The notation $\mathbf{X} \succeq 0$ indicates that matrix $\mathbf{X}$ is a positive semi-definite (PSD) matrix. $[\mathbf{X}]_{ij}$ represents $(i,j)$-th element of matrix $\mathbf{X}$.
$\mathbb{E}[\cdot]$ indicates the statistical expectation.  $\mathcal{CN}(\mu,\sigma^2)$ and $\mathcal{U}(a,b)$ represent a complex Gaussian random variable with a mean of $\mu$ and a variance $\sigma^2$ and a uniform random variable in the interval $[a,b]$, respectively.

\section{System model} \label{Sec2}
\begin{figure}[!t]
\centering
\includegraphics[width=0.9\linewidth]{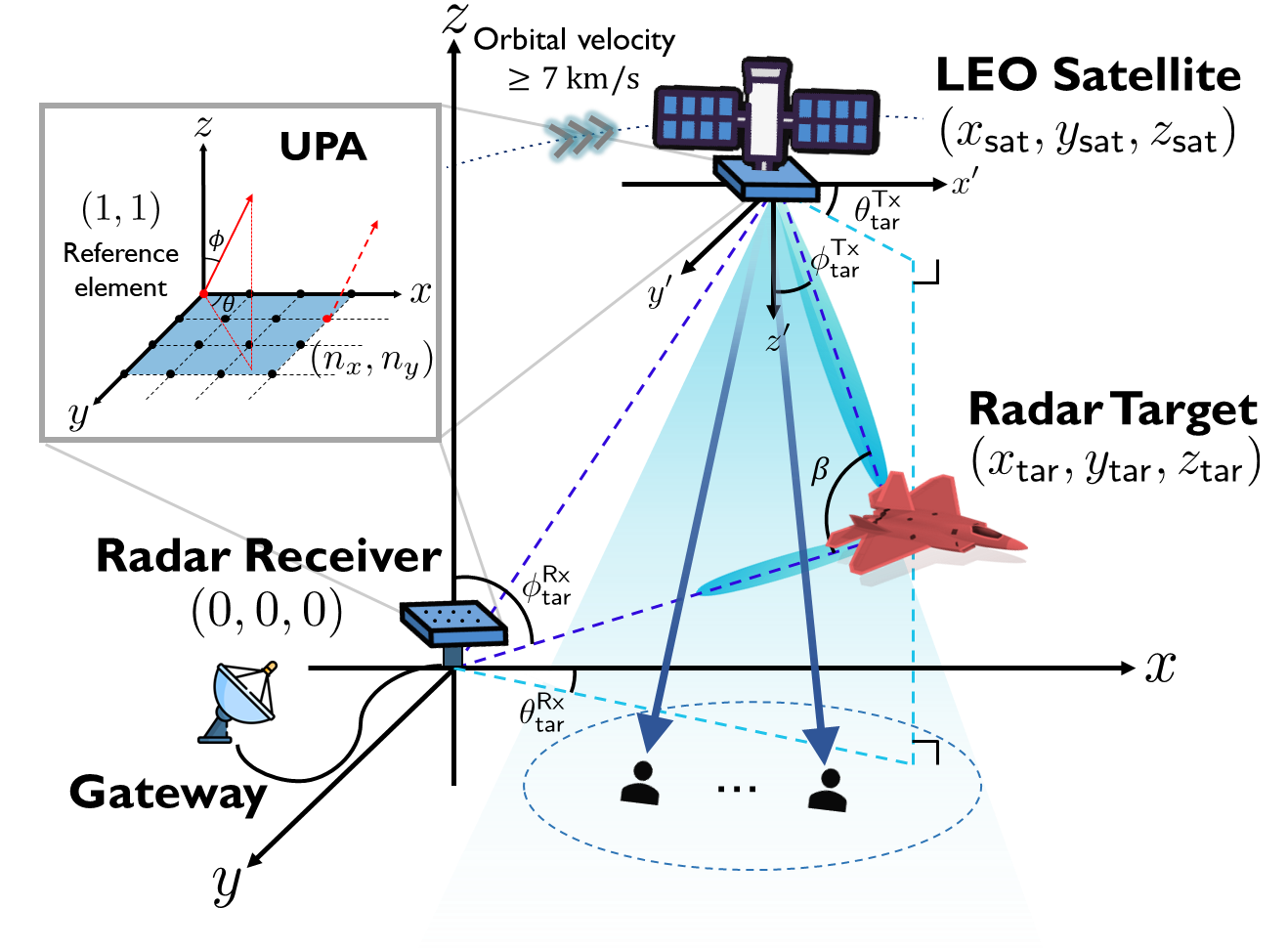}
 		\caption{The proposed bistatic LEO-ISAC system model.}
    	\label{Fig1}\vspace{0mm}
\end{figure}

We introduce a Cartesian coordinate system with the radar receiver at the origin, denoted by $(0,0,0)$. The LEO satellite is located at $(x_{\sf sat}, y_{\sf sat}, z_{\sf sat})$ as shown in Fig. \ref{Fig1}. The satellite is equipped with a uniform planar array (UPA) antenna consisting of $N_{\sf Tx}$ elements. Similarly, the radar receiver has a UPA antenna with $N_{\sf Rx}$ elements. The satellite provides communication services to $K$ single-antenna users, indexed by $\mathcal{K}\triangleq\left\{1,...,K\right\}$. $(\theta_k^{\sf Tx},\phi_k^{\sf Tx})$ stands for the AOD pair of the $k$-th user, representing the azimuth and off-nadir angles, respectively. Using the same time-frequency resources and antennas, the satellite performs radar operations to track the radar target located at $(x_{\sf tar}, y_{\sf tar}, z_{\sf tar})$ within its coverage. The corresponding AOD and AOA pairs of the radar target with respect to the satellite and the radar receiver are represented by $\left(\theta_{\sf tar}^{\sf Tx}, \phi_{\sf tar}^{\sf Tx}\right)$ and $\left(\theta_{\sf tar}^{\sf Rx}, \phi_{\sf tar}^{\sf Rx}\right)$, respectively. 
In addition, as illustrated in Fig. \ref{block_diagram}, we incorporate a link between the gateway and the radar receiver to enable cooperation between them. In particular, this link can be used for time synchronization and exchanging previously estimated target parameters and transmitted signals \cite{leyva2022two}.
\subsection{Channel Models}
\subsubsection{Satellite Communication Channel}
We employ the widely used ray-tracing approach to model the LEO satellite communication channel. Due to the high altitude of the satellite, we can assume that the AOD remains the same for each multipath \cite{you2020massive, you2022beam}. Consequently, the propagation delay for the $p$-th path between the $(n_x, n_y)$-th element of the satellite's antenna and the $k$-th user can be described as
\begin{equation}\label{propagation delay}
\tau_{k,p}^{(n_x,n_y)}=\tau_{k,p}^{\sf ref}+\Delta\tau_{k}^{(n_x,n_y)}(\theta_k^{\sf Tx},\phi_k^{\sf Tx}),
\end{equation}
where $n_x\in\left\{1,\cdots,N_{\sf Tx}^x\right\}$ and $n_y\in\left\{1,\cdots,N_{\sf Tx}^y\right\}$. Here, $N_{\sf Tx}^x$ and $N_{\sf Tx}^y$ denote the numbers of antennas on the x- and y-axes. $\tau_{k,p}^{\sf ref}\triangleq\tau_{k,p}^{(1,1)}$ represents the delay between the reference element and the $k$-th user over the $p$-th path; $\Delta\tau_{k}^{(n_x,n_y)}(\theta_k^{\sf Tx},\phi_k^{\sf Tx})$ stands for the delay between the reference element and the $(n_x,n_y)$-th element. Assuming a half-wavelength antenna spacing, $\Delta\tau_{k}^{(n_x,n_y)}(\theta_k^{\sf Tx},\phi_k^{\sf Tx})$ can be expressed as follows:
\begin{align}\label{delay within UPA}
&\Delta\tau_{k}^{(n_x,n_y)}(\theta_k^{\sf Tx},\phi_k^{\sf Tx})
\nonumber\\
&=\frac{(n_x-1)\mathrm{sin}\phi_k^{\sf Tx}\mathrm{cos}\theta_k^{\sf Tx}+(n_y-1)\mathrm{sin}\phi_k^{\sf Tx}\mathrm{sin}\theta_k^{\sf Tx}}{2f_c},
\end{align}
where $f_c$ is the carrier frequency. The noiseless signal received at the $k$-th user from the $(n_x, n_y)$-th antenna element is
\begin{align}\label{general received signal}
y_k^{(n_x,n_y)}(t)=&\sum_{p=0}^{P_k}r_{k,p}x\Big(t-\tau_{k,p}^{(n_x,n_y)}\Big)\mathrm{exp}\Big(-j2\pi{f_c}\tau_{k,p}^{(n_x,n_y)}\Big)
\nonumber
\\
&\times\mathrm{exp}\Big(j2\pi v_{k,p}t\Big),
\end{align}
where $t$ is the time index; $P_k$ represents the number of non-line-of-sight (NLOS) paths to $k$-th user; $x\left(t-\tau_{k,p}^{(n_x,n_y)}\right)$ is the transmitted signal delayed by $\tau_{k,p}^{(n_x,n_y)}$; $r_{k,p}$ is the signal amplitude on the $p$-th path. Here, $p=0$ refers to the line-of-sight (LOS) path. $v_{k,p}\triangleq v_{k,p}^{\sf ut}+v_{k,p}^{\sf sat}$ is the Doppler frequency on the $p$-th path coming from the mobility of both the satellite and the user. Due to the extremely fast speed of the satellite, the Doppler shift from the mobility of each user is negligible, i.e., $v_{k,p}^{\sf sat}\approx v_{k}^{\sf sat}, \forall p\in\left\{0,\cdots ,P_k \right\}$
\cite{you2022beam}.

Supposing that the delay spread is much smaller than the signal duration, i.e., $x\left(t-\tau_{k,p}^{(n_x,n_y)}\right)\approx x\left(t- \tau_{k,0}^{(n_x,n_y)}\right), \forall p\in\left\{1,\cdots,P_k\right\}$, the channel can be approximated as a one-tap channel. We then rewrite (\ref{general received signal}) and define the time-varying channel $h_k^{(n_x,n_y)}(t)$ as
\begin{align}
&y_k^{(n_x,n_y)}(t)\approx x\left(t-\tau_{k,0}^{(n_x,n_y)}\right)
\nonumber
\\
&\;\;\;\;\times\underbrace{\sum_{p=0}^{P_k}r_{k,p}\mathrm{exp}\left(-j2\pi{f_c}\tau_{k,p}^{(n_x,n_y)}\right)\mathrm{exp}\Big(j2\pi v_{k,p}t\Big)}_{\triangleq h_k^{(n_x,n_y)}(t)}.
\end{align}
Based on (\ref{propagation delay}), we can rewrite the channel as
\begin{align}
\label{time-varying channel}
h_k^{(n_x,n_y)}(t)=\mathrm{exp}\Big(\!-\!j2\pi\!\left\{{f_c}\tau_{k,0}^{\sf ref}-v_k^{\sf sat}t\right\}\Big)
\\
\nonumber
\times{g_k(t)a^{(n_x,n_y)}\left(\theta_k^{\sf Tx},\phi_k^{\sf Tx}\right)},
\end{align}
where complex channel coefficient $g_k(t)$ and phase shift within the antenna array $a^{(n_x,n_y)}\left(\theta_t^k,\phi_t^k\right)$ are defined as
\begin{align}
\label{channel gain}
&g_k(t)\triangleq
\sum_{p=0}^{P_k}r_{k,p}\mathrm{exp}\Big(\!-\!j2\pi\left\{f_c(\tau_{k,p}^{\sf ref}-\tau_{k,0}^{\sf ref})-v_{k,p}^{\sf ut}t\right\}\Big),
\\
\label{phase shift}
&a^{(n_x,n_y)}\left(\theta_k^{\sf Tx},\phi_k^{\sf Tx}\right)\triangleq
\mathrm{exp}\Big(\!-j2\pi{f_c}\Delta\tau^{(n_x,n_y)}\left(\theta_k^{\sf Tx},\phi_k^{\sf Tx}\right)\Big),
\end{align} respectively.
The time-varying phase shift term $\mathrm{exp}\{-j2\pi({f_c}\tau_{k,0}^{\sf ref}-v_k^{\sf sat}t)\}$ in (\ref{time-varying channel}) can be compensated with proper time-frequency synchronization \cite{you2020massive, you2022beam}. Consequently, the equivalent channel in vector form is
\begin{align}
\mathbf{h}_k(t)\triangleq 
g_k(t)\mathbf{a}\left(\theta_k^{\sf Tx},\phi_k^{\sf Tx}\right)\in\mathbb{C}^{N_{\sf Tx}\times1},
\end{align}
 where the array response vector $\mathbf{a}\left(\theta_{t}^{k},\phi_{t}^{k}\right)\in\mathbb{C}^{N_{\sf Tx}\times1}$ is the stacked version of (\ref{phase shift}), which can be represented using the Kronecker product operator as
\begin{equation}\label{steering vector}
\mathbf{a}\left(\theta,\phi\right) = \mathbf{a}_x\left(\theta,\phi\right)\otimes\mathbf{a}_y\left(\theta,\phi\right),
\end{equation}
where $\mathbf{a}_x\left(\theta,\phi\right)$ and $\mathbf{a}_y\left(\theta,\phi\right)$ are defined as follows:
\begin{align}
&\mathbf{a}_x\left(\theta,\phi\right)=\begin{bmatrix}
1,e^{-j\pi\mathrm{sin}\phi\mathrm{cos}\theta},
\cdots,  e^{-j\pi(N_{\sf Tx}^x-1)\mathrm{sin}\phi\mathrm{cos}\theta}
\end{bmatrix}^{\sf T},
\nonumber
\\
&\mathbf{a}_y\left(\theta,\phi\right)=\begin{bmatrix}
1,e^{-j\pi\mathrm{sin}\phi\mathrm{sin}\theta},
\cdots,  e^{-j\pi(N_{\sf Tx}^y-1)\mathrm{sin}\phi\mathrm{sin}\theta}
\end{bmatrix}^{\sf T}.
\nonumber
\end{align}
In the following, we focus on each coherence time interval and omit the time index $t$. We model $g_k$ as a complex random variable following a Rician distribution since the LOS path exists. Specifically, the real and imaginary parts of $g_k$ are independently and identically distributed, following $\mathcal{N}\left(\sqrt{\frac{\kappa_k\gamma_k}{2(\kappa_k+1)}},\frac{\gamma_k}{2(\kappa_k+1)}\right)$ \cite{you2020massive}. Here, $\kappa_k$ and $\gamma_k$ represent the Rician factor and the average channel power, i.e., $\gamma_k\triangleq\mathbb{E}\left[|g_k|^2\right]$. Due to the long propagation delay and short channel coherence time of LEO satellites, we assume that obtaining iCSI at the satellite is infeasible. Instead, we presume that the satellite is available to obtain both the \emph{geometrical} information of the users (i.e.,
$\{(\theta_k^{\sf Tx},\phi_k^{\sf Tx})\}_{k\in\mathcal{K}}$) and the \emph{statistical} information of the channel gain (i.e., $\{\gamma_k\}_{k\in\mathcal{K}}$) \cite{you2020massive,you2022beam}.
\subsubsection{Bistatic Radar Channel}
Assuming a point target, we model the radar channel as follows \cite{yan2007multitarget}:
\begin{align}
\mathbf{H}_{\sf R} = \alpha\mathbf{b}(\theta_{\sf tar}^{\sf Rx},\phi_{\sf tar}^{\sf Rx})\mathbf{a}^{\sf H}(\theta_{\sf tar}^{\sf Tx},\phi_{\sf tar}^{\sf Tx})\in\mathbb{C}^{N_{\sf Rx}\times N_{\sf Tx}},
\end{align}
where $\mathbf{b}(\theta_{\sf tar}^{\sf Rx},\phi_{\sf tar}^{\sf Rx})$ represents the array response vector of the radar receiver and can be modeled in the same way as in (\ref{steering vector}). $\alpha$ represents the reflection coefficient, where the power $|\alpha|^2$ is modeled based on the bistatic radar equation \cite{yan2007multitarget}, given by
\begin{align}
\label{reflection coefficient power}
\vert\alpha\vert^2 = \frac{G_{\sf sat}G_{\sf R}c^2\sigma_{\sf RCS}^{\sf bi}}{(4\pi)^3 R_{\sf Tx}^2 R_{\sf Rx}^2 f_c^2},
\end{align}
where $G_{\sf sat}$ and $G_{\sf R}$ are the antenna gains of the satellite and the radar receiver, respectively.\footnote{In the monostatic ISAC structure, $\sigma_{\sf RCS}^{\sf bi}$ and $R_{\sf Rx}$ are replaced by $\sigma_{\sf RCS}^{\sf mono}$ and $R_{\sf Tx}$, respectively, resulting in $\vert\alpha\vert^2\propto R_{\sf Tx}^{-4}$. Roughly speaking, when the radar target is within the coverage area, $R_{\sf Rx}$ would generally be much smaller than $R_{\sf Tx}$. This leads to a significant difference in echo path loss between monostatic and bistatic structures. A numerical analysis incorporating the angle-dependent reflection is presented in Sec. V.} $R_{\sf Tx}$ and $R_{\sf Rx}$ represent the distance from the satellite to the target and the distance from the target to the radar receiver, respectively. $c$ represents the speed of light; $\sigma_{\sf RCS}^{\sf bi}$ is a bistatic radar cross section (RCS) that can be modeled as $\sigma_{\sf RCS}^{\sf bi}=\sigma_{\sf RCS}^{\sf mono}\mathrm{cos}(\beta/2)$ \cite{kell1965derivation}, where $\sigma_{\sf RCS}^{\sf mono}$ represents the monostatic RCS; $\beta$ is a bistatic angle (defined as the angle subtended between the satellite, target, and radar receiver) as shown in Fig. 2. Note that our bistatic RCS model is a function of $\beta$; thus, we account for the angle-dependent reflection effect that is important in bistatic ISAC.
\subsection{Signal Models}
\subsubsection{Dual-Functional Transmit Signal}
Since we consider the 1-layer downlink RSMA, the messages for all users $\left\{M_1,\cdots,M_K\right\}$ are split into common parts and private parts \cite{park2023rate,9831440}. The private parts $\left\{M_{{\sf p},1},\cdots,M_{{\sf p},K}\right\}$ are independently encoded into private streams $\left\{s_1,\cdots,s_K\right\}$ using a codebook known only to the corresponding users. On the other hand, all common parts $\left\{M_{{\sf c},1},\cdots,M_{{\sf c},K}\right\}$ are combined and jointly encoded into a common stream $s_{\sf c}$ using a codebook that is known to all users. Let $\mathbf{s}[\ell] = \begin{bmatrix}s_{1}[\ell],\cdots,s_{K}[\ell],s_{\sf c}[\ell],s_{\sf R}[\ell]\end{bmatrix}^{\sf T} \in \mathbb{C}^{(K+2)\times 1}$ represents the baseband signal vector before precoding, where $s_{\sf R}[\ell]$ is a radar sequence and $\ell\in\left\{1,\cdots,L\right\}$ represents the discrete time sample index. Here, we assume that $\frac{1}{L}\sum_{\ell=1}^{L}\mathbf{s}[\ell]\mathbf{s}^{\sf H}[\ell] = \mathbf{I}_{K+2}$ asymptotically holds as $L\rightarrow \infty$. Note that $\mathbf{s}_{\sf R}$ is a pre-determined sequence that can be known \emph{a priori} by both the satellite and the users; thus, it can be used for inter-function interference management. In addition, for notational brevity, we define sets of indexes as $\mathcal{C}\triangleq\left\{1,... ,K,{\sf c}\right\}$ and $\mathcal{R}\triangleq\left\{1,...,K,{\sf c}, {\sf R}\right\}$, so $\mathcal{K}\subset\mathcal{C}\subset\mathcal{R}$ holds.

Consequently, the transmit signal at the satellite is given by
\begin{align}\label{transmit signal}
{\mathbf{x}}[\ell] &= \mathbf{P}\mathbf{s}[\ell] = \sum_{j\in\mathcal{R}}\mathbf{p}_{j}s_{j}[\ell]
\nonumber\\
&= \underbrace{\mathbf{p}_{\sf R}s_{\sf R}[\ell]}_{\rm{radar~sequence}} + \underbrace{\overbrace{\mathbf{p}_{\sf c}s_{\sf c}[\ell]}^{\rm{common~stream}}+\overbrace{\sum_{k\in\mathcal{K}}\mathbf{p}_{k}s_{k}[\ell]}^{\rm{private~streams}}}_{\rm{communication~streams}},
\end{align}
where $\mathbf{P}=\begin{bmatrix}\mathbf{p}_1,\cdots,\mathbf{p}_K,\mathbf{p}_{\sf c},\mathbf{p}_{\sf R}\end{bmatrix}\in\mathbb{C}^{N_{\sf Tx}\times{(K+2)}}$ is the dual-functional precoder matrix to be designed. Specifically, $\big\{\mathbf{p}_k\big\}_{k\in\mathcal{K}}$ represent precoding vectors corresponding to $K$ private streams, while $\mathbf{p}_{\sf c}$ and $\mathbf{p}_{\sf R}$ correspond to the common stream and radar sequence, respectively.

\subsubsection{Received Signal at Communication User}
Based on (\ref{transmit signal}), the received signal at the $k$-th user can be expressed as follows:
\begin{align}
\label{kth user received signal}
\!\!{y_k}[\ell]&={{{\mathbf{h}}^{\sf H}_k}}{\mathbf{x}[\ell]}+{z_{k}[\ell]
}= \mathbf{h}_k^{\sf H}\sum_{j\in\mathcal{R}}\mathbf{p}_{j}s_{j}[\ell]+z_{k}[\ell]
\nonumber\\
&={{\mathbf{h}}^{\sf H}_k}\bigg({\mathbf{p}_{\sf R}s_{\sf R}[\ell]}+{\mathbf{p}_{\sf c}s_{\sf c}[\ell]}+\sum_{k\in\mathcal{K}}\mathbf{p}_{k}s_k[\ell]\bigg)+{z_{k}[\ell]
},
\end{align}
where $z_{k}[\ell]\sim\mathcal{CN}(\mathbf{0},\sigma_{c}^2)$ represents complex Gaussian noise. Each user first decodes the common stream by treating other streams as interference. Therefore, the achievable common rate of the $k$-th user is
\begin{equation}\label{Common rate}
\hspace{-3mm}{R}_{c,k}=\!\log_{2}\bigg(\!1+\!\frac{|{\mathbf{h}}^{\sf H}_k\mathbf{p}_{\sf c}|^2}{\sum_{j\in\mathcal{K}}|{\mathbf{h}}^{\sf H}_k\mathbf{p}_j|^2+\delta_{\sf SIC}|{\mathbf{h}}^{\sf H}_k\mathbf{p}_{\sf R}|^2+\sigma_\mathrm{c}^2}\bigg),\!
\end{equation}
where the binary variable $\delta_{\sf SIC}$ is determined by the user's ability to perform successive interference cancellation (SIC) of the radar sequence \cite{xu2021rate}. If the SIC of $s_{\sf R}$ is available, $\delta_{\sf SIC}=0$; otherwise, $\delta_{\sf SIC}=1$. While SIC operation provides interference control capability, this leads to increased receiver complexity.

After re-encoding and precoding the common stream, it is subtracted from the received signal. Each user then decodes the private stream. Therefore, the achievable private rate of the $k$-th user can be expressed as
\begin{equation}\label{Private rate}
\hspace{-2mm}{R}_{p,k}=\log_{2}\Bigg(\!1+\frac{|{\mathbf{h}}^{\sf H}_k\mathbf{p}_k|^2}{{\sum_{\substack{j\in\mathcal{K} \\j\neq k}}}|{\mathbf{h}}^{\sf H}_k\mathbf{p}_j|^2+\delta_{\sf SIC}|{\mathbf{h}}^{\sf H}_k\mathbf{p}_{\sf R}|^2+\sigma_\mathrm{c}^2}\Bigg).
\end{equation}

Finally, the total achievable rate of the $k$-th user is given by $R_{p,k} + C_k$, where $C_k$ represents the portion of the common rate dedicated to the $k$-th user. The common rate allocation $\{C_k\}_{k\in\mathcal{K}}$ will be jointly optimized with the precoder $\mathbf{P}$ in the following section. It is essential to ensure that every user can decode the common stream. Therefore, a constraint is imposed on the sum of the common rate portions, i.e., $\sum_{k\in\mathcal{K}}C_k\leq\ \min_{k}R_{c,k}$. 
The achievable rate expressions (\ref{Common rate}) and (\ref{Private rate}) depend on iCSI, which is not suitable for designing precoders for LEO satellites. Motivated from \cite{you2022beam}, we define the upper bound of the ergodic rate as follows:
\begin{align}\label{common_bound}
\overline{{R}}_{c,k}&\triangleq\!\log_{2}\bigg(\!1\!+\frac{\mathbb{E}\left[|{\mathbf{h}}^{\sf H}_k\mathbf{p}_{\sf c}|^2\right]}{\mathbb{E}\big[\sum_{j\in\mathcal{K}}|{\mathbf{h}}^{\sf H}_k\mathbf{p}_j|^2+\delta_{\sf SIC}|{\mathbf{h}}^{\sf H}_k\mathbf{p}_{\sf R}|^2\big]+\sigma_\mathrm{c}^2}\bigg)
\nonumber\\
&=\!\log_{2}\bigg(\!1\!+\frac{\gamma_k\vert\mathbf{a}_k^{\sf H}\mathbf{p}_{\sf c}\vert^2}{\gamma_k\sum_{j\in\mathcal{K}}\vert\mathbf{a}_k^{\sf H}\mathbf{p}_j\vert^2+\gamma_k\delta_{\sf SIC}\vert\mathbf{a}_k^{\sf H}\mathbf{p}_{\sf R}\vert^2+\sigma_\mathrm{c}^2}\bigg)
\nonumber\\
&=\!\log_{2}\bigg(\!1\!+\frac{\vert\mathbf{a}_k^{\sf H}\mathbf{p}_{\sf c}\vert^2}{\sum_{j\in\mathcal{K}} \vert\mathbf{a}_k^{\sf H}\mathbf{p}_j\vert^2+\delta_{\sf SIC}\vert\mathbf{a}_k^{\sf H}\mathbf{p}_{\sf R}\vert^2+\rho_k}\bigg),\!
\end{align}
\begin{align}\label{private_bound}
\overline{{R}}_{p,k}&\triangleq\!\log_{2}\Bigg(\!1\!+\frac{\mathbb{E}\left[|{\mathbf{h}}^{\sf H}_k\mathbf{p}_k|^2\right]}{\mathbb{E}\Big[\sum_{\substack{j\in\mathcal{K} \\j\neq k}}|{\mathbf{h}}^{\sf H}_k\mathbf{p}_j|^2+\delta_{\sf SIC}|{\mathbf{h}}^{\sf H}_k\mathbf{p}_{\sf R}|^2\Big]+\sigma_\mathrm{c}^2}\Bigg)
\nonumber\\&=\!
\log_{2}\Bigg(\!1\!+\frac{\gamma_k\vert\mathbf{a}_k^{\sf H}\mathbf{p}_k\vert^2}{\gamma_k\sum_{\substack{j\in\mathcal{K} \\j\neq k}}\vert\mathbf{a}_k^{\sf H}\mathbf{p}_j\vert^2+\gamma_k\delta_{\sf SIC}\vert\mathbf{a}_k^{\sf H}\mathbf{p}_{\sf R}\vert^2+\sigma_\mathrm{c}^2}\Bigg)
\nonumber\\&=\!
\log_{2}\Bigg(\!1\!+\frac{\vert\mathbf{a}_k^{\sf H}\mathbf{p}_k\vert^2}{\sum_{\substack{j\in\mathcal{K} \\j\neq k}}\vert\mathbf{a}_k^{\sf H}\mathbf{p}_j\vert^2+\delta_{\sf SIC}\vert\mathbf{a}_k^{\sf H}\mathbf{p}_{\sf R}\vert^2+\rho_k}\Bigg),\!\!
\end{align}
where $\rho_k\triangleq\ \sigma_c^2 \slash \gamma_k$ represents the noise power-to-average channel power ratio for the $k$-th user; $\mathbf{a}_k\triangleq\mathbf{a}(\theta_k^{\sf Tx},\phi_k^{\sf Tx})$ represents the array response corresponding to the AOD of the $k$-th user. Here, $\overline{R}_{{\sf c},k}\geq\mathbb{E}\left[{R}_{{\sf c},k}\right]$ and $\overline{R}_{p,k}\geq\mathbb{E}\left[{R}_{p,k}\right]$ holds. The proof is a straightforward application of Jensen's inequality based on the fact that ($\ref{Common rate}$) and ($\ref{Private rate}$) are concave functions with respect to $|g_k|^2$ \cite{you2022beam}. Hence, we omitted it here for brevity. Note that $(\ref{common_bound})$ and $(\ref{private_bound})$ are expressed in terms of the geometrical information (i.e., $\{\mathbf{a}_k\}_{k\in\mathcal{K}}$) and statistical information (i.e., $\{\rho_k\}_{k\in\mathcal{K}}$) of the communication channels. In contrast to iCSI, this information is often readily available or can be easily determined thanks to the quasi-deterministic nature of satellite movement following a predetermined orbit around the Earth \cite{roper2022beamspace}. Furthermore, this information exhibits much slower variations than iCSI, making it suitable for the precoder design of LEO satellite systems.
\\
\subsubsection{Received Echo Signal at Radar Receiver}
Let $\mathbf{S}\triangleq\begin{bmatrix}\mathbf{s}[1],\mathbf{s}[2],\cdots,\mathbf{s}[L]\end{bmatrix}\in\mathbb{C}^{(K+2)\times L}$ and $\mathbf{X}\triangleq\begin{bmatrix}\mathbf{x}[1],\mathbf{x}[2],\cdots,\mathbf{x}[L]\end{bmatrix}\in\mathbb{C}^{N_{\sf Tx}\times L}$ denote the stacked versions of $\mathbf{s}[\ell]$ and $\mathbf{x}[\ell]$, respectively. 
The received echo signal reflected from the radar target can be expressed as
\begin{equation}\label{received reflected signal, bistatic}
\mathbf{Y}_{\sf R}=\mathbf{H}_{\sf R}\tilde{\mathbf{X}}+\mathbf{Z}_{\sf R},
\end{equation}
where $\mathbf{Z}_{\sf R}$ is a noise matrix with each element following independent and identically distributed circularly symmetric complex Gaussian noise with zero mean and a variance of $\sigma_{\sf R}^2$. The matrix $\tilde{\mathbf{X}}\in\mathbb{C}^{N_{\sf Tx}\times (L+\tau_{\sf max})}$ represents a delayed and Doppler-shifted version of $\mathbf{X}$, which can be modeled as
\begin{equation}
\tilde{\mathbf{X}}=\begin{bmatrix}
\mathbf{X}\mathbf{V}(v_{\sf tar}),\mathbf{0}_{N_{\sf Tx}\times\tau_{\sf max}}
\end{bmatrix}\mathbf{J}(\tau_{\sf tar}),
\end{equation}
where $\mathbf{V}(v_{\sf tar})\triangleq\mathrm{diag}\left(e^{j2\pi v_{\sf tar}T_s},\cdots,e^{j2\pi v_{\sf tar}LT_s}\right)$ is a diagonal matrix that represents the Doppler effect with a Doppler frequency of $v_{\sf tar}$ and a sampling interval of $T_s$. $\mathbf{J}(\tau_{\sf tar})$ is a time-domain shifting matrix, which has the form \cite{liu2020joint}
\begin{equation}
    \mathbf{J}(\tau_{\sf tar}) = \begin{bmatrix}
    \smash[b]{\begin{matrix}0 & \cdots & 0 \end{matrix}} &1 &\cdots &0 \\
    \smash[b]{\begin{matrix}0& \cdots& 0\end{matrix}} & 0 & \ddots & \vdots \\
        \vdots & \vdots & \vdots & 1 \\
            \vdots & \vdots &\vdots &\vdots
            \\
            \smash[b]{\underbrace{\begin{matrix}0 & \cdots & 0 \end{matrix}}_{{\tau_{\sf tar}}\text{ zeros}}} &0 &\cdots &0
    \end{bmatrix}, \tau_{\sf tar}\in\left\{1, \cdots, \tau_{\sf max}\right\},
\end{equation}
\\
where $\tau_{\sf tar}$ is the delayed number of samples; $\tau_{\sf max}$ is the maximum delayed number of samples. 
\begin{remark}
{\rm \textbf{(Relationship between proposed bistatic ISAC and conventional monostatic ISAC):}} When the position of the satellite and the radar receiver are identical, i.e., the origin is set as the satellite, our system model boils down to the monostatic ISAC structure. 
While in this work, we assume that the radar receiver is separated from the satellite, it is evident that our proposed bistatic ISAC can be considered as a unified one in that it encompasses the monostatic LEO-ISAC (such as system models in \cite{you2022beam} and \cite{liu2024max}) as a special case.
\end{remark}
\begin{remark}{\rm \textbf{(Robustness to uplink-echo interference):}} Unlike in a monostatic ISAC, in the proposed system model, the satellite receives the uplink communication signals, while the radar receiver receives the radar echo signals. 
Therefore, interference between the uplink and echo signals is naturally avoided regardless of the duplex scheme \cite{pucci2022performance}.
\end{remark}
\begin{remark}
{\rm \textbf{(Extension to satellite-based radar receivers):}}
Although this work focuses on the scenario where the radar receiver is on the ground, in cases where the goal is to detect high-altitude radar targets such as missiles or satellites, the radar receiver could be another LEO satellite. In this scenario, an inter-satellite link can facilitate cooperation between the two satellites 
 by \tcb{enabling} time synchronization and exchange of useful information \cite{kodheli2020satellite}.
\end{remark}
\section{Precoder Optimization for Bistatic LEO-ISAC}\label{Sec3}
To design an appropriate dual-functional precoder, we use the CRB as a radar performance metric, which is a lower bound on the variance of any unbiased estimator. Note that CRB is known to be tight to the mean squared error (MSE) in the high signal-to-noise (SNR) region \cite{sun2023trade}. To derive the CRB, we begin with the truncated version of (\ref{received reflected signal, bistatic}) as follows:\footnote{The Doppler effect is neglected here because it does not affect the expression of the CRB of AOA estimation. This is evident from the fact that $\mathbf{V}_v\mathbf{V}_v^{\sf H}=\mathbf{I}_L, \forall v\in\mathbb{R}$ and the derivation of (\ref{F_theta_theta}).}
\begin{equation}\label{echo signal without delay doppler}
\!\!\!\!\Acute{\mathbf{Y}}_{\sf R}=\mathbf{H}_{\sf R}\mathbf{X}+\Acute{\mathbf{Z}}_{\sf R}=\alpha\mathbf{b}(\theta_{\sf tar}^{\sf Rx},\phi_{\sf tar}^{\sf Rx})\mathbf{a}^{\sf H}(\theta_{\sf tar}^{\sf Tx},\phi_{\sf tar}^{\sf Tx}){\mathbf{X}}+\Acute{\mathbf{Z}}_{\sf R},
\end{equation}
where $\Acute{\mathbf{Z}}_\mathrm{R}$ is the clipped version of $\mathbf{Z}_\mathrm{R}$ in (\ref{received reflected signal, bistatic}). Hence, $\rm{vec}(\Acute{\mathbf{Y}}_{\sf R})$ is complex Gaussian random vector that follows  $\mathcal{CN}\left(\boldsymbol{\mu}, \boldsymbol{\Sigma}\right)$, where $\boldsymbol{\mu}=\alpha
\mathrm{vec}\left(\mathbf{b}(\theta_{\sf tar}^{\sf Rx},\phi_{\sf tar}^{\sf Rx})\mathbf{a}^{\sf H}(\theta_{\sf tar}^{\sf Tx},\phi_{\sf tar}^{\sf Tx})\mathbf{X}\right)$ and $\boldsymbol{\Sigma}=\sigma_{\sf R}^2\mathbf{I}_{LN_{\sf Rx}}$.
Let us define the vector of unknown AOA of echo signal as $\boldsymbol{\xi}=[\theta_{\sf tar}^{\sf Rx},\phi_{\sf tar}^{\sf Rx}]^{\sf T}$. Subsequently, the Fisher information matrix (FIM) can be calculated as \cite{kay1993fundamentals}
\begin{equation}\label{Fisher information matrix}
\left[\mathbf{F}\right]_{ij}=\underbrace{\mathrm{tr}
\left(
\boldsymbol{\Sigma}^{-1}\frac{\partial\boldsymbol{\Sigma}}{\partial{{\xi}}_i}
\boldsymbol{\Sigma}^{-1}\frac{\partial\boldsymbol{\Sigma}}{\partial{{\xi}}_j}
\right)}_{=0}
+
2\mathrm{Re}\left(
\frac{\partial\boldsymbol{\mu}^{\sf H} }{\partial{{\xi}}_i}\boldsymbol{\Sigma}^{-1}\frac{\partial\boldsymbol{\mu}}{\partial{{\xi}}_j}
\right),
\end{equation}
where ${\xi}_i$ represent the $i$-th element of the $\boldsymbol{\xi}$. In (\ref{Fisher information matrix}), the first term is zero since the noise covariance matrix is independent of AOA. We express FIM for notational simplicity as
\begin{equation}\label{block FIM}
\mathbf{F}=\begin{bmatrix} 
F_{\theta\theta} & F_{\theta\phi} \\ F_{\phi\theta} & F_{\phi\phi} 
\end{bmatrix}.
\end{equation}
Let us denote $\mathbf{a}_{\sf tar}\triangleq\mathbf{a}(\theta_{\sf tar}^{\sf Tx}, \phi_{\sf tar}^{\sf Tx})$ and define the partial derivatives as $\dot{\mathbf{b}}_{\theta}\triangleq\frac{\partial}{\partial\theta_{\sf tar}^{\sf Rx}}\mathbf{b}(\theta_{\sf tar}^{\sf Rx},\phi_{\sf tar}^{\sf Rx})$ and $\dot{\mathbf{b}}_{\phi}\triangleq\frac{\partial}{\partial\phi_{\sf tar}^{\sf Rx}}\mathbf{b}(\theta_{\sf tar}^{\sf Rx},\phi_{\sf tar}^{\sf Rx})$. Consequently, the first element of $\mathbf{F}$ can be calculated as
\begin{align}
\label{F_theta_theta}
&F_{\theta\theta}\nonumber
\nonumber\\
&=2\mathrm{Re}\left\{\vert\alpha\vert^2\mathrm{vec}\left(\dot{\mathbf{b}}_{\theta}\mathbf{a}_{\sf tar}^{\sf H}\mathbf{X}\right)^{\sf H}\frac{1}{\sigma_\mathrm{R}^2}\mathbf{I}_{LN_{\sf Rx}}\mathrm{vec}\left(\dot{\mathbf{b}}_{\theta}\mathbf{a}_{\sf tar}^{\sf H}\mathbf{X}\right)\right\}
\nonumber\\
&\overset{(\rm{a})}{=}\frac{2\vert\alpha\vert^2}{\sigma_\mathrm{R}^2}\mathrm{Re}\left\{\mathrm{tr}\left(\dot{\mathbf{b}}_{\theta}\mathbf{a}_{\sf tar}^{\sf H}\mathbf{X}\mathbf{X}^{\sf H}\mathbf{a}_{\sf tar}\dot{\mathbf{b}}^{\sf H}_{\theta}\right)\right\}
\nonumber\\
&\overset{(\rm{b})}{=}\frac{2L\vert\alpha\vert^2}{\sigma_\mathrm{R}^2}\mathrm{tr}\left(\mathbf{a}_{\sf tar}^{\sf H}\mathbf{P}\mathbf{P}^{\sf H}\mathbf{a}_{\sf tar}\dot{\mathbf{b}}^{\sf H}_{\theta}\dot{\mathbf{b}}_{\theta}\right)
\nonumber\\&=\frac{2L\vert\alpha\vert^2}{\sigma_{\sf R}^2}\Vert\mathbf{P}^{\sf H}\mathbf{a}_{\sf tar}\Vert_2^2\Vert\dot{\mathbf{b}}_{\theta}\Vert_2^2.
\end{align}
 In (a), we use the fact that $\mathrm{vec}(\mathbf{A})^{\sf H}\mathrm{vec}(\mathbf{B}) = \mathrm{tr}(\mathbf{B}\mathbf{A}^{\sf H})$. We utilize the property that eigenvalues of a Hermitian matrix are real, and the assumption $\mathbf{SS}^{\sf H}=\sum_{\ell=1}^{L}\mathbf{s}[\ell]\mathbf{s}^{\sf H}[\ell]=L\mathbf{I}_{K+2}$ asymptotically holds as $L\rightarrow \infty$ for (b). In the same manner, we can obtain the remaining elements as follows:
\begin{align}
&F_{\theta\phi}
=F_{\phi_\theta}=\frac{2L\vert\alpha\vert^2}{\sigma_{\sf R}^2}\Vert\mathbf{P}^{\sf H}\mathbf{a}_{\sf tar}\Vert_2^2\mathrm{Re}\big(\dot{\mathbf{b}}^{\sf H}_{\theta}\dot{\mathbf{b}}_{\phi}\big),
\nonumber
\\
&F_{\phi\phi}
=\frac{2L\vert\alpha\vert^2}{\sigma_{\sf R}^2}\Vert\mathbf{P}^{\sf H}\mathbf{a}_{\sf tar}\Vert_2^2\Vert\dot{\mathbf{b}}_{\phi}\Vert^2_2.
\end{align}
The CRB can be obtained as diagonal elements of $\mathbf{F}^{-1}$ as
\begin{equation}
\label{CRB_theta}
{\sf CRB}_{\theta}=\frac{Q\Vert\dot{\mathbf{b}}_{\phi}\Vert_2^2}{\Vert\mathbf{P}^{\sf H}\mathbf{a}_{\sf tar}\Vert_2^2}, \quad{\sf CRB}_{\phi}= \frac{Q\Vert\dot{\mathbf{b}}_{\theta}\Vert_2^2}{\Vert\mathbf{P}^{\sf H}\mathbf{a}_{\sf tar}\Vert_2^2},
\end{equation} where $Q=\sigma_{\sf R}^2 \slash \big\{2L|\alpha|^2(\Vert\dot{\mathbf{b}}_{\theta}\Vert_2^2\Vert\dot{\mathbf{b}}_{\phi}\Vert_2^2-\mathrm{Re}(\dot{\mathbf{b}}^{\sf H}_{\theta}\dot{\mathbf{b}}_{\phi})^2)\big\}$. Here, the denominator of (\ref{CRB_theta}) represents the beamforming gain corresponding to the radar target. Thus, constraining the CRB of the AOA is equivalent to ensuring that a certain amount of power is \tcb{directed} toward the radar target. This in turn leads to improved estimation performance not only for AOA but also for delay and Doppler.

Finally, our problem to jointly optimize the dual-functional precoder $\mathbf{P}$ and the common rate portions $\{C_k\}_{k\in\mathcal{K}}$ that maximize the minimum rate among all users while satisfying a given level of CRB  can be formulated as follows:
\begin{maxi!}|l|[2]
{_{\mathbf{P},\left\{{C_k}\right\}_{k \in \mathcal{K}}}}
{\min_k\left(\overline{R}_{p,k}+C_k\right)\label{objective_function}}
{\label{optimizationProblem}}
{}
\addConstraint{{\sf CRB}_{\theta}\leq\gamma_{\theta}^{\sf th}, {\sf CRB}_{\phi}\leq\gamma_{\phi}^{\sf th},}\label{radar CRB constraint}
\addConstraint{\min_k\left(\overline{R}_{c,k}\right)\geq \sum_{k\in\mathcal{K}}C_k,}\label{common-decoding}
\addConstraint{C_k \geq 0, \forall k\in\mathcal{K},}\label{non-negative-common}
\addConstraint{\mathrm{tr}\big(\mathbf{PP}^{\sf H}\big)}{\leq P_t.\label{power_const}}
\end{maxi!}
Here, (\ref{radar CRB constraint}) represents the CRB constraints, where $\gamma_{\theta}^{\sf th}$ and $\gamma_{\phi}^{\sf th}$ represent thresholds associated with elevation and off-nadir angle, respectively. (\ref{common-decoding}) represents the necessary condition for the common stream to be successfully decoded by all users. (\ref{non-negative-common}) is introduced to prohibit the common rate portions from having negative values; (\ref{power_const}) is a power budget constraint.

We first introduce an auxiliary variable $R_{\sf min}$ and reformulate the epigraph form of (\ref{optimizationProblem}), given by
\begin{maxi!}|l|[2]
{_{\mathbf{P},\left\{{C_k}\right\}_{k \in \mathcal{K}},{R_{\sf min}}}}
{R_{\sf min}}
{\label{epigraph_form}}
{}
\addConstraint{\overline{R}_{p,k}+C_k\geq R_{\sf min}, \forall k\in\mathcal{K},}\label{min_rate_const}
\addConstraint{\overline{R}_{c,k}\geq \sum_{{\tcb{k'}}\in\mathcal{K}}C_{\tcb{k'}}, \forall k\in\mathcal{K},}\label{common-decoding 2}
\addConstraint{\textrm{(\ref{radar CRB constraint}), (\ref{non-negative-common}), (\ref{power_const}).}}\nonumber
\end{maxi!}
To address the non-convex problem (\ref{epigraph_form}) and develop an efficient optimization algorithm, we employ various relaxation techniques. For enhanced readability, we divide the procedures and present them as follows:

\subsection*{{\bf Step I) Semidefinite Relaxation}}
Since the CRB constraint (\ref{radar CRB constraint}) is non-convex, we employ the SDR technique \cite{luo2010semidefinite}. Let us define rank-1 matrices as $\overline{\mathbf{P}}_j\triangleq \mathbf{p}_j\mathbf{p}_j^{\sf H},$ $ \forall j\in \mathcal{R}$, $\mathbf{A}_{\sf tar}\triangleq\mathbf{a}_{\sf tar}\mathbf{a}_{\sf tar}^{\sf H}$, and $\mathbf{A}_k\triangleq\mathbf{a}_k\mathbf{a}_k^{\sf H}, \forall k\in\mathcal{K}.$ Consequently, the CRB constraint can be reformulated by using the properties $\Vert\mathbf{P}^{\sf H}\mathbf{a}_{\sf tar}\Vert_2^2$ $=\mathbf{a}_{\sf tar}^{\sf H}\mathbf{PP}^{\sf H}\mathbf{a}_{\sf tar}$ $=\mathrm{tr}\left(\mathbf{A}_{\sf tar}\mathbf{PP}^{\sf H}\right)$ and $\mathbf{PP}^{\sf H}=\sum_{j\in\mathcal{R}}\overline{\mathbf{P}}_j$ as
\begin{align}
\label{CRB_const_afterSDR}
&\mathrm{tr}\left(\mathbf{A}_{\sf tar}\sum_{j\in\mathcal{R}}\overline{\mathbf{P}}_j\right)\geq\rm{max}\left\{\frac{Q\Vert\dot{\mathbf{b}}_{\phi}\Vert_2^2}{\gamma_{\theta}^{\sf th}},\frac{Q\Vert\dot{\mathbf{b}}_{\theta}\Vert_2^2}{\gamma_{\phi}^{\sf th}}\right\}.
\end{align}

Similarly, we rewrite $\overline{R}_{{\sf c},k}$ and $\overline{R}_{p,k}$ by using $\vert\mathbf{a}_k^{\sf H}\mathbf{p}_j\vert^2$ $= \mathbf{a}_k^{\sf H}\mathbf{p}_j\mathbf{p}_j^{\sf H}\mathbf{a}_k$ $=\mathrm{tr}(\mathbf{a}_k\mathbf{a}_k^{\sf H}\mathbf{p}_j\mathbf{p}_j^{\sf H})$ $= \mathrm{tr}\left(\mathbf{A}_k\overline{\mathbf{P}}_j\right)$ as
\begin{equation}
\label{Common rate after SDR}
\overline{R}_{{\sf c},k}
=\log_{2}\bigg(\frac{\sum_{j\in\mathcal{C}}\mathrm{tr}\left({\mathbf{A}}_k\overline{\mathbf{P}}_j\right)+\delta_{\sf SIC}\mathrm{tr}\left({\mathbf{A}}_k\overline{\mathbf{P}}_{\sf R}\right)+\rho_k}{\sum_{j\in\mathcal{K}}\mathrm{tr}\left({\mathbf{A}}_k\overline{\mathbf{P}}_j\right)+\delta_{\sf SIC}\mathrm{tr}\left({\mathbf{A}}_k\overline{\mathbf{P}}_{\sf R}\right)+\rho_k}\bigg)
\end{equation}
and
\begin{equation}
\label{Private rate after SDR}
\overline{R}_{p,k}
=\log_{2}\Bigg(\frac{\sum_{j\in\mathcal{K}}\mathrm{tr}\left({\mathbf{A}}_k\overline{\mathbf{P}}_j\right)+\delta_{\sf SIC}\mathrm{tr}\left({\mathbf{A}}_k\overline{\mathbf{P}}_{\sf R}\right)+\rho_k}{\sum_{\substack{j\in\mathcal{K} \\j\neq k}}\mathrm{tr}\left({\mathbf{A}}_k\overline{\mathbf{P}}_j\right)+\delta_{\sf SIC}\mathrm{tr}\left({\mathbf{A}}_k\overline{\mathbf{P}}_{\sf R}\right)+\rho_k}\Bigg),
\end{equation}
respectively. 

\subsection*{{\bf Step II) Sequential Rank-1 Constraint Relaxation}}

By employing the SDR technique, it leads to the emergence of both the PSD constraints and the rank-1 constraints for $\{\overline{\mathbf{P}}_{j}\}_{j\in\mathcal{R}}$. While the PSD constraints are convex, the rank-1 constraints pose a challenge. To address the non-convexity of rank-1 constraints, we employ the SROCR technique instead of simply dropping the rank-1 constraint \cite{cao2017sequential}. Based on the fact that the trace of a matrix is equal to the sum of its eigenvalues, we can reformulate the rank-1 constraints equivalently as
\begin{align}
\lambda_{\sf max}\left(\overline{\mathbf{P}}_j\right)=\mathrm{tr}\left(\overline{\mathbf{P}}_j\right), \forall j\in\mathcal{R},
\end{align}
where $\lambda_{\sf max}\left(\cdot\right)$ denote the maximum eigenvalue. By defining $\mathbf{v}_{j, \sf max}$ as the principal eigenvector corresponding to $\overline{\mathbf{P}}_{j}$, we can rewrite the equivalent rank-1 constraint as
\begin{align}
\label{reformulated rank-1 const}
\mathbf{v}_{j, \sf max}^{\sf H}\overline{\mathbf{P}}_{j}\mathbf{v}_{j, \sf max} = \mathrm{tr}\left(\overline{\mathbf{P}}_{j}\right), \forall j\in\mathcal{R}.
\end{align} Since (\ref{reformulated rank-1 const}) is still non-convex with respect to $\overline{\mathbf{P}}_{j}$, we relax it to make it more tractable as follows:
\begin{align} \label{relexed rank-1 constraint}
&\mathbf{v}_{j, \sf max}^{[m]\sf H}\overline{\mathbf{P}}_{j}\mathbf{v}^{[m]}_{j, \sf max}\geq w_{j}^{[m]}\mathrm{tr}\left(\overline{\mathbf{P}}_{j}\right), \forall j\in \mathcal{R},
\end{align}
where $m$ represents the iteration index of the outer layer, and accordingly, $\mathbf{v}^{[m]}_{j, \sf max}$ is the principal eigenvector corresponding to the solution matrix $\overline{\mathbf{P}}^{[m-1]}_{j}$ obtained from the $[m-1]$-th iteration. $w_{j}^{[m]}\in[0,1]$ serves as relaxation parameters, controlling the ratio of the maximum eigenvalue to the trace. By monotonically increasing values of $\{w_{j}^{[m]}\}_{j\in\mathcal{R}}$, we can systematically force the solution matrices to become close to rank-1. By doing so, we effectively mitigate the loss of optimality caused by using SDR. 


\subsection*{{\bf Step III) Introducing Auxiliary Variables and Successive Convex Approximation}}
To tackle the remaining non-convexity of our optimization problem, we derive the upper and lower bounds of the denominators and numerators in (\ref{Common rate after SDR}) and (\ref{Private rate after SDR}) by introducing auxiliary variables $\mathbf{c,d,e,f} \in \mathbb{R}^{K\times1}$ as
\begin{align}\label{ak}
&\sum_{j\in\mathcal{C}}\mathrm{tr}\left({\mathbf{A}}_k\overline{\mathbf{P}}_j\right)+\delta_{\sf SIC}\mathrm{tr}\left({\mathbf{A}}_k\overline{\mathbf{P}}_{\sf R}\right)+\rho_k \geq e^{c_k}, \forall k\in\mathcal{K},
\\
\label{nonconvex1}
&\sum_{j\in\mathcal{K}}\mathrm{tr}\left({\mathbf{A}}_k\overline{\mathbf{P}}_j\right)+\delta_{\sf SIC}\mathrm{tr}\left({\mathbf{A}}_k\overline{\mathbf{P}}_{\sf R}\right)+\rho_k\leq e^{d_k}, \forall k\in\mathcal{K},
\\
\label{ck}
&\sum_{j\in\mathcal{K}}\mathrm{tr}\left({\mathbf{A}}_k\overline{\mathbf{P}}_j\right)+\delta_{\sf SIC}\mathrm{tr}\left({\mathbf{A}}_k\overline{\mathbf{P}}_{\sf R}\right)+\rho_k\geq e^{e_k}, \forall k\in\mathcal{K},
\\
\label{nonconvex2}
&\sum_{\substack{j\in\mathcal{K} \\j\neq k}}\mathrm{tr}\left({\mathbf{A}}_k\overline{\mathbf{P}}_j\right)+\delta_{\sf SIC}\mathrm{tr}\left({\mathbf{A}}_k\overline{\mathbf{P}}_{\sf R}\right)+\rho_k\leq e^{f_k}, \forall k\in\mathcal{K},
\end{align}
respectively. Here, $c_k$ denotes the $k$-th element of vector $\mathbf{c}$, and the same applies to $\mathbf{d}$, $\mathbf{e}$, and $\mathbf{f}$.
By utilizing these bounds, we can establish the lower bounds of $\overline{R}_{{\sf c},k}$ and $\overline{R}_{p,k}$ as follows:
\begin{align}
\overline{R}_{{\sf c},k}\geq\frac{1}{\mathrm{ln}2}\left(c_k-d_k\right), 
\overline{R}_{p,k}\geq\frac{1}{\mathrm{ln}2}\left(e_k-f_k\right), \forall k\in\mathcal{K}.
\end{align}
By doing so, we can modify (\ref{min_rate_const}) and (\ref{common-decoding 2}) as follows:
\begin{align} \label{min-rate const bound}
&\frac{1}{\mathrm{ln}2}\left(e_k-f_k\right)+C_k\geq R_{\sf min}, \forall k \in\mathcal{K},
\\ \label{common-decoding const bound}
&\frac{1}{\mathrm{ln}2}\left(c_k-d_k\right)\geq\sum_{{\tcb{k'}}\in\mathcal{K}}C_{\tcb{k'}}, \forall k \in\mathcal{K},
\end{align} respectively. Since (\ref{nonconvex1}) and (\ref{nonconvex2}) are still non-convex, we utilize the SCA method. Using the first-order Taylor series approximation, the exponential terms in (\ref{nonconvex1}) and (\ref{nonconvex2}) can be approximated, given by
\begin{align}\label{bk}
&\sum_{j\in\mathcal{K}}\mathrm{tr}\left({\mathbf{A}}_k\overline{\mathbf{P}}_j\right)+\delta_{\sf SIC}\mathrm{tr}\left({\mathbf{A}}_k\overline{\mathbf{P}}_{\sf R}\right)+\rho_k\
\nonumber\\
&\qquad\qquad\qquad\leq e^{d^{[n-1]}_k}\left(d_k-d_k^{[n-1]}+1\right), \forall k \in\mathcal{K},
\\
\label{dk}
&\hspace{-3mm}\sum_{{j\in\mathcal{K}, j\neq k}}\mathrm{tr}\left({\mathbf{A}}_k\overline{\mathbf{P}}_j\right)+\delta_{\sf SIC}\mathrm{tr}\left({\mathbf{A}}_k\overline{\mathbf{P}}_{\sf R}\right)+\rho_k
\nonumber\\
&\qquad\qquad\qquad\leq e^{f^{[n-1]}_k}\left(f_k-f_k^{[n-1]}+1\right), \forall k \in\mathcal{K},
\end{align}
where $n$ is the iteration index of the inner layer. Note that our algorithm turns out to be a two-layer iterative algorithm by employing both SROCR and SCA algorithms. The relaxed version of the original problem can be represented as the $[m, n]$-th subproblem, which is given by
\begin{maxi!}|l|[2]
{\substack{\left\{{\overline{\mathbf{P}}}_{j}\right\}_{j\in\mathcal{R}},\left\{{C_k}\right\}_{k\in\mathcal{K}},\\{R_{\sf min}},\mathbf{c},\mathbf{d},\mathbf{e},\mathbf{f}}}
{R_{\sf min}\label{objective}}
{\label{opt_final}}
{}
\addConstraint{\mathrm{tr}\left(\sum_{j\in\mathcal{R}}\overline{\mathbf{P}}_j\right)\leq P_t,}
\addConstraint{\overline{\mathbf{P}}_{j}\succeq\mathbf{0}, \forall j \in\mathcal{R},}\label{PSD const}
\addConstraint{\textrm{(\ref{non-negative-common}), (\ref{CRB_const_afterSDR}), (\ref{relexed rank-1 constraint})},}\nonumber
\addConstraint{\textrm{(\ref{ak}), (\ref{ck}), \eqref{min-rate const bound}--\eqref{dk}}}.\nonumber
\end{maxi!}
Since problem (\ref{opt_final}) is convex, it can be efficiently solved using off-the-shelf optimizers, such as CVX \cite{grant2014cvx}. We summarize the process of precoder optimization in Algorithm 1.\footnote{In practice, the AOD of the radar target $(\theta_k^{\sf Tx},\phi_k^{\sf Tx})$, an input parameter of Algorithm 1, should be the predicted AOD rather than the true values. For instance, this predicted AOD can be the output of extended Kalman filter \cite{kulikov2015accurate}.}

In Algorithm 1, we begin by initializing $\{w^{[0]}_j\}_{j\in\mathcal{R}}=0$, which is equivalent to dropping rank-1 constraints. We then proceed to the inner layer as lines 5-9 to find solutions that are possibly high-rank matrices. Next, we determine $\big\{w^{[1]}_j\big\}_{j\in\mathcal{R}}$ based on the maximum eigenvalue-to-trace ratio of the obtained solution matrices (as line 15), and return to the inner layer. Herein, due to the relaxed rank-1 constraints (\ref{relexed rank-1 constraint}), the subproblem in the subsequent inner layer is possibly unsolvable. In that case, we reduce the stepsize (as line 12) by half and update $\big\{w^{[m]}_j\big\}_{j\in\mathcal{R}}$ (as lines 12-15), and return to the inner layer. By repeating the same steps, the algorithm terminates and obtains nearly rank-1 solution matrices when $\big\{w_j^{[m]}\big\}_{j\in\mathcal{R}}$ converge to 1. Finally, the optimized precoder $\mathbf{P}^{\star}$ can be obtained by applying eigenvalue decomposition to $\big\{\overline{\mathbf{P}}_j^{\star}\big\}_{j\in\mathcal{R}}$ \cite{luo2010semidefinite}.
\vspace{-3mm}
\begin{algorithm}[!t]
\caption{Two-Layer Iterative Precoder Optimization Algorithm for Proposed Bistatic LEO-ISAC}\label{Algorithm 1}
\begin{algorithmic}[1]
\State \textbf{Input}: $\alpha, \theta_{\sf tar}^{\sf Tx}, \phi_{\sf tar}^{\sf Tx}, \gamma_\theta^{\sf th}, \gamma_\phi^{\sf th}, 
\gamma_k,
\theta_k^{\sf Tx}, \phi_k^{\sf Tx}, \delta_{\sf SIC}, P_{t}, \forall k\in\mathcal{K}$
\State \textbf{Initialize}: $m \leftarrow 0, n \leftarrow 0, \{w_j^{[0]}\}_{j\in\mathcal{R}} \leftarrow 0, \delta^{[0]}, \mathbf{d}^{[0]}, \mathbf{f}^{[0]}$
\While{$\exists w_j^{[m]} {\rm{~s.t.~}} w_j^{[m]}<\epsilon_{\sf rank}, \forall j\in\mathcal{R}$ \& $m<m_{\sf max}$}
\If{Problem (\ref{opt_final}) is \emph{solvable}}
\While{$\vert R_{\sf min}^{[n]}-R_{\sf min}^{[n-1]}\vert>\epsilon_{\sf obj}$ \& $n< n_{\sf max}$}
\State Solve (\ref{opt_final}) to obtain $\mathbf{d}^{[n]}, \mathbf{f}^{[n]}$
      \State $\mathbf{d}^{[n+1]} \leftarrow \mathbf{d}^{[n]}, \mathbf{f}^{[n+1]} \leftarrow \mathbf{f}^{[n]}$
      \State $n \leftarrow n+1$
    \EndWhile
    \State $\delta^{[m+1]}\leftarrow\delta^{[0]}$
\Else
\State 
$\delta^{[m+1]}\leftarrow\frac{1}{2}\delta^{[m]}$
\State $\overline{\mathbf{P}}_j^{[m]}\leftarrow\overline{\mathbf{P}}_j^{[m-1]},\forall j\in\mathcal{R}$
\EndIf
    \State $w_{j}^{[m+1]}\leftarrow\mathrm{min}\left(1, \frac{\lambda_{\sf max}\left(\overline{\mathbf{P}}_{j}^{[m]}\right)}{\mathrm{tr}\left(\overline{\mathbf{P}}_{j}^{[m]}\right)}+\delta^{[m+1]}\right), \forall j\in\mathcal{R}$
    \State $m\leftarrow m+1, n\leftarrow 0$
\EndWhile
\State \textbf{Output}: $\big\{\overline{\mathbf{P}}_j^{\star}\big\}_{j\in\mathcal{R}}, \big\{C_k^{\star}\big\}_{k\in\mathcal{K}}$ 
\end{algorithmic}
\end{algorithm}
\begin{figure}[!t]
\centering
 		\includegraphics[width=0.85\linewidth]{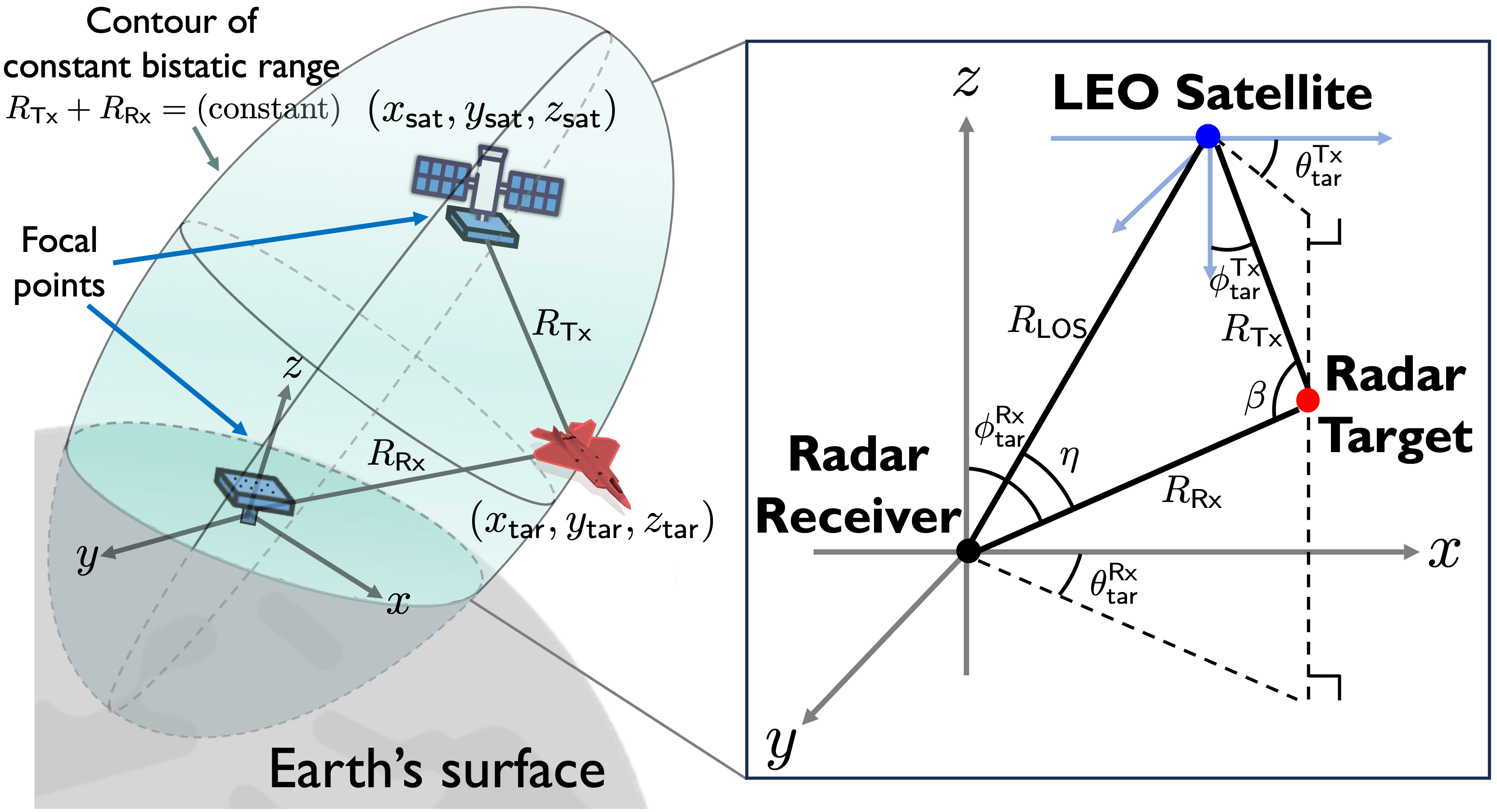}
 		\caption{Geometrical relationship among the radar receiver, LEO satellite, and the radar target in the proposed framework.}
    	\label{geometry}\vspace{0mm}
\end{figure}
\section{Target Parameter Estimation for Radar} \label{Sec4}
In this section, we present the radar target parameter estimation process that can be employed in our proposed framework. These include the conventional MUSIC algorithm and a novel joint delay-Doppler-AOD estimation algorithm that exploits the geometry of the bistatic structure.\footnote{As the radar works to track the target, we focus on parameter estimation. In practice, periodic detection of the target would be required.}
\subsection*{{\bf Step I) Multiple Signal Classification}}
To estimate the AOA of the echo signal, the conventional MUSIC algorithm can be used \cite{MUSIC}.
From (\ref{received reflected signal, bistatic}), the sample covariance matrix can be computed and rewritten using the eigenvalue decomposition theorem as follows:
\begin{equation}
\mathbf{R_\mathrm{Y}} = \frac{1}{L+\tau_{\sf max}}\mathbf{Y}_{\sf R}\mathbf{Y}_{\sf R}^{\sf H}=
\begin{bmatrix}
\mathbf{v}_{\sf s},\mathbf{V}_{\sf n}
\end{bmatrix}
\begin{bmatrix}
\sigma_{\sf s}  &  \mathbf{0}^{\sf T}\\
\mathbf{0} &  \Sigma_{\sf n}
\end{bmatrix}
\begin{bmatrix}
\mathbf{v}^{\sf H}_{\sf s} \\ \mathbf{V}^{\sf H}_{\sf n}
\end{bmatrix},
\end{equation}
where $\mathbf{v}_{\sf s}\in\mathbb{C}^{N_{\sf Rx}\times{1}}$ and column vectors in $\mathbf{V}_{\sf n}\in\mathbb{C}^{N_{\sf Rx}\times{(N_{\sf Rx}-1)}}$ span the signal and noise subspaces, respectively. The AOA of the radar target can be estimated as
\begin{equation}\label{DOA estimation}
\left(
\hat{\theta}_{\sf tar}^{\sf Rx},\hat{\phi}_{\sf tar}^{\sf Rx}\right)
=\arg\max_{\theta,\phi}\frac{1}{\mathbf{b}(\theta,\phi)^{\sf H}{\mathbf{U}}_n{\mathbf{U}}_n^{\sf H}\mathbf{b}(\theta,\phi)}.
\end{equation}
\subsection*{{\bf Step II) Joint Delay-Doppler-AOD Estimation}}
We put forth a matched filter algorithm for jointly estimating delay, Doppler, and AOD. By leveraging the previously estimated AOA, the known satellite positions, and the geometry of the bistatic structure, we can compute the AOD for each delay candidate in the searching space. This is possible because the position of the target can be uniquely determined for a given delay and the AOA. Specifically, for a given delay, the contour of a constant bistatic range has the shape of an ellipsoid, with the satellite and radar receiver located at the focal points \cite{marom2022bistatic}. 
Here, the bistatic range refers to the sum of the radar receiver-to-target and target-to-satellite distances, i.e., $R_{\sf Rx}+R_{\sf Tx}$ in Fig. 3. Then, the intersection of the line in the direction of AOA with this ellipsoid uniquely determines the position of the target, thereby retrieving AOD from the known satellite position. By doing so, an appropriate reference signal can be generated for a given delay and Doppler candidates in the searching space. This can be seen as a modified and extended version of the algorithm proposed in \cite{liu2020joint} to 3-dimensional bistatic geometry. In the following, we present the mathematical details of the proposed algorithm. We first present the detailed procedure for computing the AOD for a given delay candidate. Firstly, we define the unit direction vectors of the satellite and the AOA of echo signal with respect to the radar receiver as
\begin{equation}
\mathbf{d}_{\sf sat}=\frac{
\begin{bmatrix}
x_{\sf sat}, y_{\sf sat}, z_{\sf sat}
\end{bmatrix}^{\sf T}}
{\left\|\begin{bmatrix}
x_{\sf sat}, y_{\sf sat}, z_{\sf sat}
\end{bmatrix}\right\|_2},
\end{equation}
\begin{equation}
\label{AOA vector}
\begin{split}
\hat{\mathbf{d}}_{\sf AOA}=
\left[\mathrm{sin}\hat{\phi}_{\sf tar}^{\sf Rx}\mathrm{cos}\hat{\theta}_{\sf tar}^{\sf Rx}\right.,
\mathrm{sin}\hat{\phi}_{\sf tar}^{\sf Rx}\mathrm{sin}\hat{\theta}_{\sf tar}^{\sf Rx},
\left.\mathrm{cos}\hat{\phi}_{\sf tar}^{\sf Rx}
\right]^{\sf T},\end{split}
\end{equation}respectively. Then, the angle $\eta$ subtended between the satellite, radar receiver, and radar target (as shown in Fig. 3) can be obtained as
\begin{equation}
\hat{\eta}=\mathrm{cos}^{-1}\big(\mathbf{d}_{\sf sat}^{\sf T}\hat{\mathbf{d}}_{\sf AOA}\big).
\end{equation}
Using a given $\tau$, we can calculate the bistatic range as
\begin{equation}
R_{\sf Rx}+R_{\sf Tx}=c\times\tau T_s,
\end{equation}
where $T_s$ is the sampling interval. So far we know the bistatic range, but $R_{\sf Rx}$ is unknown. By using cosine law, $R_{\sf Rx}$ can be obtained as follows:
\begin{equation}
R_{\sf Rx}=\frac{\left(R_{\sf Rx}+R_{\sf Tx}\right)^2-R_{\sf LOS}^2}{2\left(R_{\sf Rx}+R_{\sf Tx}-R_{\sf LOS}\times\mathrm{cos}\hat{\eta}\right)},
\end{equation}
where $R_{\sf LOS}$ represents the radar receiver-to-satellite distance. Hence, the position of the (potential) target is
\begin{equation}
\begin{bmatrix}
{x}_{\sf tar}, {y}_{\sf tar}, {z}_{\sf tar}
\end{bmatrix}^{\sf T}
=
R_{\sf Rx}\times\hat{\mathbf{d}}_{{\sf AOA}}.
\end{equation}
Finally, we can compute the AOD for a given $\tau$ as follows:
\begin{align}
&\theta_{\sf tar}^{\sf Tx}=
\begin{cases}
    \mathrm{tan}^{-1}\left(\frac{y_{\sf tar}-y_{\sf sat}}{x_{\sf tar}-x_{\sf sat}}\right)& \text{if } x_{\sf tar}>x_{\sf sat},\\
    \mathrm{tan}^{-1}\left(\frac{y_{\sf tar}-y_{\sf sat}}{x_{\sf tar}-x_{\sf sat}}\right)+\pi& \text{if } x_{\sf tar}<x_{\sf sat}, y_{\sf tar}\geq y_{\sf sat},
    \\
        \mathrm{tan}^{-1}\left(\frac{y_{\sf tar}-y_{\sf sat}}{x_{\sf tar}-x_{\sf sat}}\right)-\pi& \text{if } x_{\sf tar}<x_{\sf sat}, y_{\sf tar}<y_{\sf sat},
        \\
        +\frac{\pi}{2} & \text{if } x_{\sf tar}=x_{\sf sat}, y_{\sf tar}>y_{\sf sat},
        \\
                -\frac{\pi}{2} & \text{if } x_{\sf tar}=x_{\sf sat}, y_{\sf tar}<y_{\sf sat},
\end{cases}
\\
&\phi_{\sf tar}^{\sf Tx}=\mathrm{cos}^{-1}\left(\frac{z_{\sf sat}-z_{\sf tar}}{\left\|[x_{\sf tar}-x_{\sf sat}, y_{\sf tar}-y_{\sf sat}, z_{\sf tar}-z_{\sf sat}]\right\|_2}\right).
\end{align}
This approach allows us to generate a reference signal for each delay and Doppler frequency candidate as
\begin{align}\label{reference signal}
\mathbf{Y}_{\sf R}^{\sf ref}=\mathbf{b}(\hat{\theta}_{\sf tar}^{\sf Rx},\hat{\phi}_{\sf tar}^{\sf Rx})\mathbf{a}^{\sf H}\left(\theta_{\sf tar}^{\sf Tx},\phi_{\sf tar}^{\sf Tx}\right)\begin{bmatrix}\mathbf{X}\mathbf{V}(v),\mathbf{0}_{N_{\sf Rx}\times \tau_{\sf max}}\end{bmatrix}\mathbf{J}(\tau).
\end{align}
However, using (\ref{reference signal}) directly as the reference signal is computationally inefficient for the matched filter algorithm, which requires a large number of computations of the correlation between the received signal and the reference signal. Therefore, we use the signal after receive combining $\mathbf{y}_{\sf R}^{\sf H}\triangleq\mathbf{b}(\hat{\theta}_{\sf tar}^{\sf Rx},\hat{\phi}_{\sf tar}^{\sf Rx})^{\sf H}\mathbf{y}_{\sf R}$, and can generate the reference signal as 
\begin{align}\label{reference signal after receive beamforming}
\left(\mathbf{y}_{\sf R}^{\sf ref}\right)^{\sf H}=\mathbf{a}^{\sf H}\left(\theta_{\sf tar}^{\sf Tx},\phi_{\sf tar}^{\sf Tx}\right)\begin{bmatrix}\mathbf{X}\mathbf{V}(v),\mathbf{0}_{N_{\sf Rx}\times \tau_{\sf max}}\end{bmatrix}\mathbf{J}(\tau).
\end{align}
This approach not only reduces the computational burden but also helps to cancel unwanted interference, such as clutter and jamming signals, when accurate AOA estimation is achieved. Specific details of the proposed algorithm are described in Algorithm 2.
 Through such a process, we can determine the position of the radar target.

 \begin{algorithm}[!t]
\caption{Matched Filter Algorithm for Joint Delay-Doppler-AOD Estimation}\label{Algorithm 2}
\begin{algorithmic}[1]
\State \textbf{Input}: $\mathbf{X}, \mathbf{Y}_{\sf R}, \hat{\theta}_{\sf tar}^{\sf Rx}, \hat{\phi}_{\sf tar}^{\sf Rx}$
\For{all $\tau$ in searching space}
      \State Compute $\theta_{\sf tar}^{\sf Tx}, \phi_{\sf tar}^{\sf Tx}$ based on $\hat{\theta}_{\sf tar}^{\sf Rx}, \hat{\phi}_{\sf tar}^{\sf Rx}$, $\tau$
      \For{all $v$ in searching space}
      \State Generate reference signal as in (\ref{reference signal after receive beamforming})
      \State Compute $\left|\left(\mathbf{y}_{\sf R}^{\sf H}\mathbf{y}_{\sf R}^{\sf ref}\right)\right|$
\EndFor
\EndFor
\State \textbf{Output}:
$(\hat{\tau}, \hat{v})=\arg\max_{\tau,v}\left|\left(\mathbf{y}_{\sf R}^{\sf H}\mathbf{y}_{\sf R}^{\sf ref}\right)\right|$; $(\hat\theta_{\sf tar}^{\sf Tx}, \hat\phi_{\sf tar}^{\sf Tx})$ can be obtained from $\hat{\tau}$.
\end{algorithmic}
\end{algorithm}


\section{Simulation Results} \label{Sec5}
In this section, we present the simulation results. We set the average power of the communication channel as
\begin{equation}
\gamma_k=G_{\sf sat}G_{\sf ut}\left(\frac{c}{4\pi f_c d}\right)^2,
\end{equation}
where $G_{\sf ut}$ and $d$ represent the antenna gain of the communication user and the distance between the satellite and the user, respectively. The noise power is defined as $\sigma_{\sf R}^2 = \sigma_{c}^2 = k_{\sf B}BT_{\sf n}$ with the Boltzmann constant $k_{\sf B} = 1.38\times10^{-23} \mathrm{J\cdot K^{-1}}$. The communication users are uniform-randomly distributed on the coverage with a diameter of $\num{100}$ km, and we set $L = \num{4096}$ samples. The remaining simulation parameters are listed in Table I.
\begin{table}[!t]
\centering
    \caption{Simulation Parameters}
    \renewcommand{\arraystretch}{1.5}
    \begin{tabular}{P{0.1\linewidth}P{0.17\linewidth}P{0.262\linewidth}P{0.273\linewidth}}
    \Xhline{1.5pt} 
      \cellcolor{White}& \textbf{Abbreviation} & \textbf{Definition} & \textbf{Value}\\
      \Xhline{1.5pt} 
      \cellcolor{White}& $f_c$ & Carrier frequency & $\num{2}$ GHz\\
      \rowcolor{Gray}
      \cellcolor{White}& $c$ & Speed of light & $\num{3}\times10^8$ m/s\\
      \cellcolor{White}& $\kappa_k$ & Rician factor & $\num{10}$ dB\\
      \rowcolor{Gray}
      \multirow{-4}{*}{\cellcolor{White}\textbf{Channel}} & $B$ & Bandwidth & $\num{10}$ MHz\\
      \hline
      \cellcolor{White}& $G_{\sf sat}$ & Antenna gain & $\num{6}$ dBi\\
      \rowcolor{Gray}
      \cellcolor{White}&$N_{\sf Tx}$ & Number of antennas & $\num{64}~(\num{8}\times\num{8}~{\rm UPA})$\\ 
       \multirow{-3}{*}{\cellcolor{White}\textbf{Satellite}} & \!\!\!$(x_{\sf sat},y_{\sf sat},z_{\sf sat})$ & Position & $(\num{30}, \num{-30}, \num{340})$ km\\
      \hline
      \rowcolor{Gray}
      \cellcolor{White} &$G_{\sf R}$ & Antenna gain & $\num{3}$ dBi\\
      \cellcolor{White}&$N_{\sf Rx}$ & Number of antennas & $\!\!\num{1024}~(\num{32}\times\num{32}~{\rm UPA})$\\
      \rowcolor{Gray}
      \cellcolor{White}&$T_{\sf n}$ & Noise temperature & $\num{150}$ K\\
      \multirow{-4}{*}{\cellcolor{White}\shortstack{{\bf Radar}\\{\bf receiver}}} & $-$ & Position & $(\num{0}, \num{0}, \num{0})$ km\\
      \hline
      \rowcolor{Gray}
      \cellcolor{White}&$G_{\sf ut}$ & Antenna gain & $\num{0}$ dBi\\
      \cellcolor{White}&$K$ & Number of users & $\num{4}$\\
      \rowcolor{Gray}
      \multirow{-3}{*}{\cellcolor{White}\shortstack{{\bf Comm.}\\{\bf user}}} &$T_{\sf n}$ & Noise temperature & $\num{150}$ K\\
      \hline
      \cellcolor{White}& $\sigma_{\sf RCS}^{\sf mono}$ & Monostatic RCS & $\num{100}$ $\rm m^2$\\
      \rowcolor{Gray}
      \cellcolor{White}& $v_{\sf tar}$ & Doppler frequency & $\sim\mathcal{U}(\num{-30},\num{30})$ kHz\\
      \multirow{-3}{*}{\cellcolor{White}\shortstack{{\bf Radar}\\{\bf target}}} & \!\!\!$(x_{\sf tar},y_{\sf tar},z_{\sf tar})$ & Position & $(\num{3}, \num{3}, \num{5})$ km\\
      \Xhline{1.5pt} 
      \end{tabular}
      \renewcommand{\arraystretch}{1}
\end{table}
\begin{figure}[!t]
\centering
 	\includegraphics[width=0.9\linewidth]{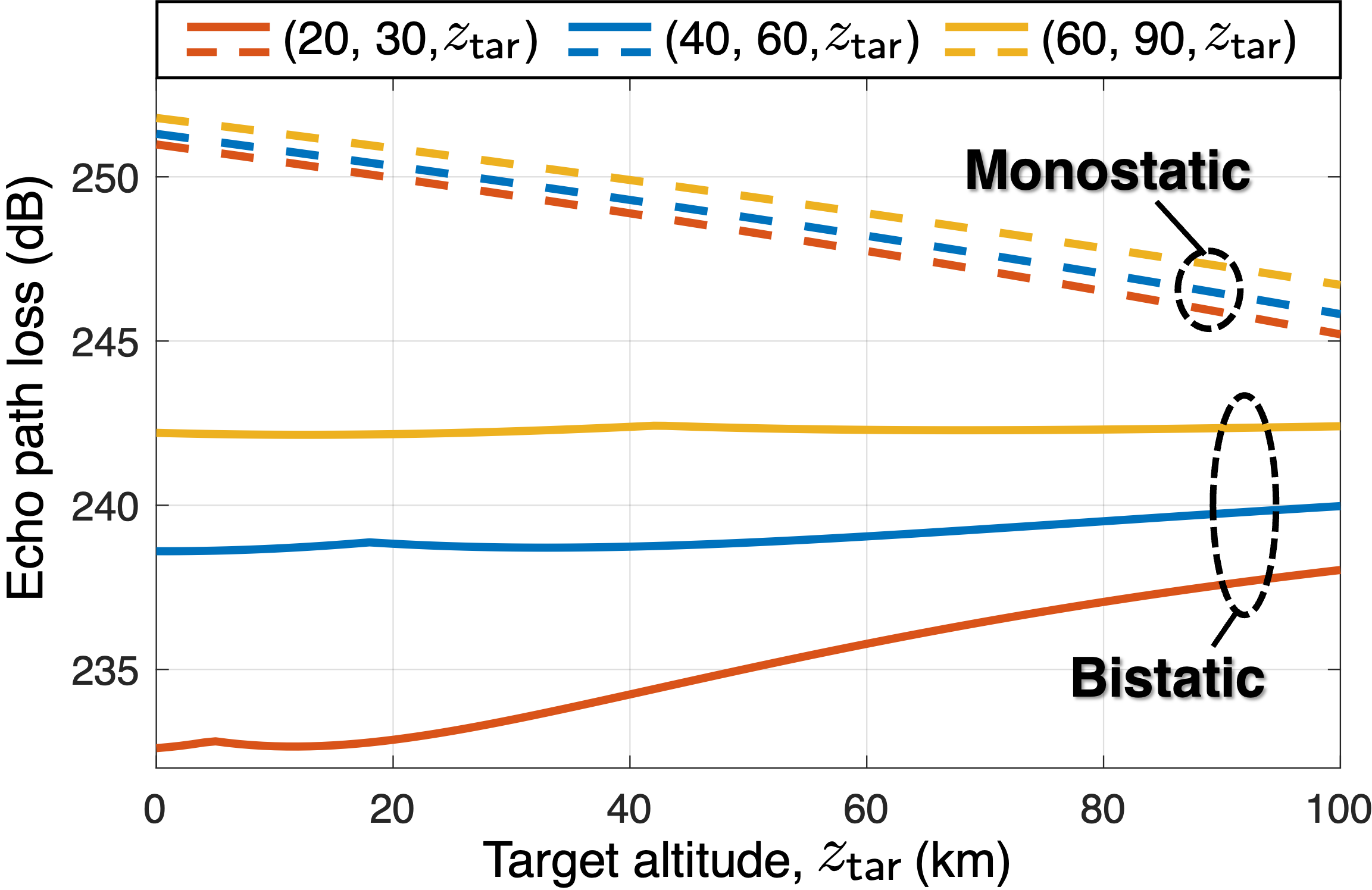}
 		\caption{Echo path loss with increasing target altitude in the bistatic and monostatic structures at the different target's $x$-$y$ coordinates.}
    	\label{mono_vs_bi}\vspace{0mm}
\end{figure}
\subsection{Bistatic vs. Monostatic Comparison}
We evaluate the bistatic and monostatic structures by comparing the echo path loss for a fair comparison. Fig. \ref{mono_vs_bi} shows the echo path loss with increasing altitude at different $x$ and $y$ coordinates of the target. The echo path loss here includes not only the round-trip free-space loss but also the effects of target reflection as in (\ref{reflection coefficient power}). It is evident from the figure that the bistatic structure exhibits significantly lower loss compared to the monostatic configuration. Note that in our RCS model, power reflection is strongest at the angle of incidence, decreasing with the angle $\beta$. However, the numerical results indicate that the impact of free space loss is more dominant. Furthermore, the gap between the two structures widens as the target altitude decreases. This highlights the benefit of the bistatic structure, particularly when considering altitudes below 20 km, which is relevant for real-world targets such as drones, aircraft, and high-altitude platforms (HAPs).

\begin{figure}[!t]
\centering
 	\includegraphics[width=0.9\linewidth]{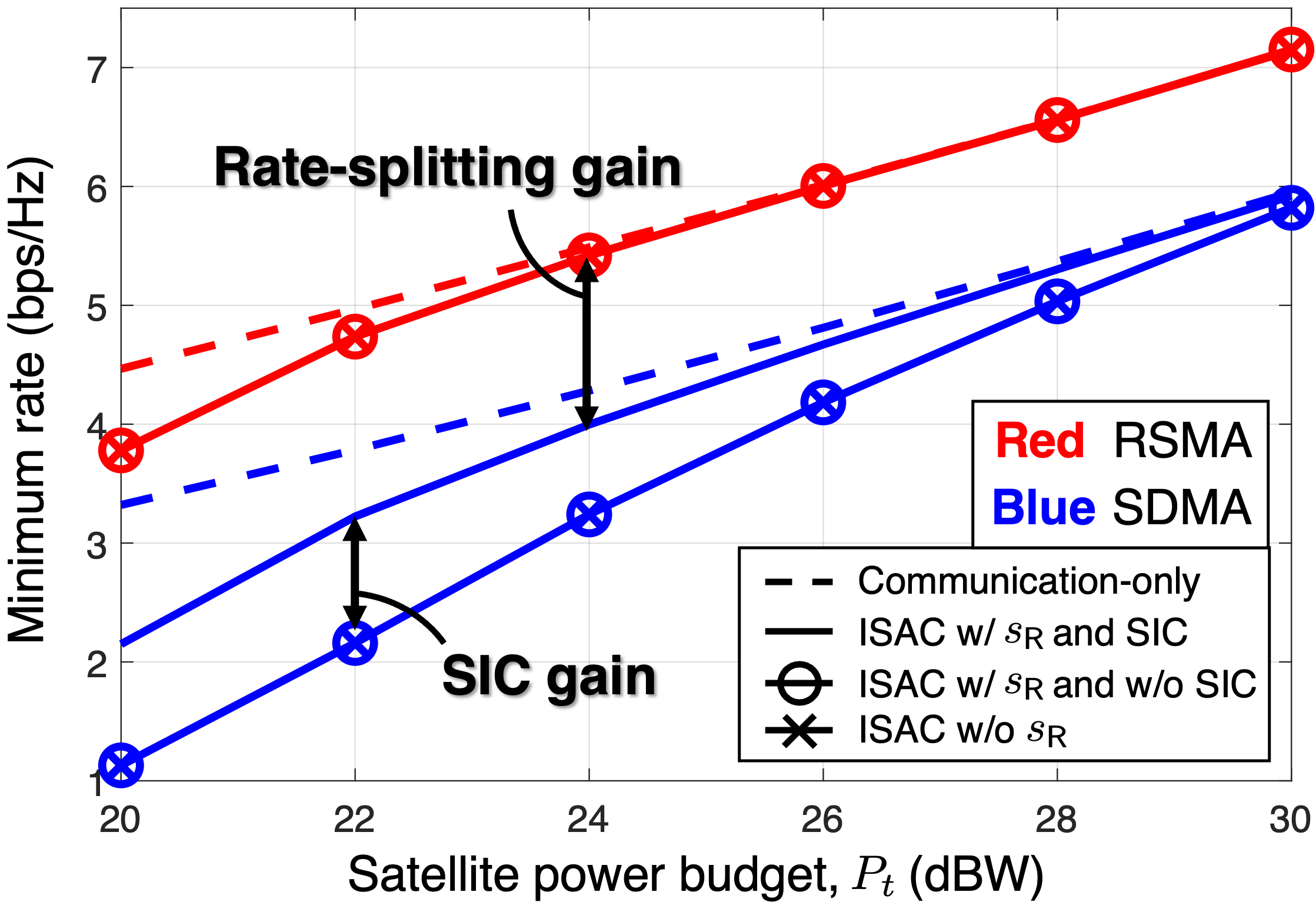}
 		\caption{Minimum rate performance under the various transmission modes with the CRB thresholds $\gamma_\theta^{\sf th}=\gamma_\phi^{\sf th}=8\times10^{-7}~{\rm rad}$.}
    	\label{Communication_Performance}\vspace{-3mm}
\end{figure}
\begin{table}[!t]
\centering
    \caption{Variable Settings for Various Transmission Modes.}
    \renewcommand{\arraystretch}{1.7}
    \begin{tabular}{P{0.27\linewidth}P{0.29\linewidth}P{0.29\linewidth}}
    \toprule[1.5pt]
      & \textbf{RSMA} & \textbf{SDMA}\\
      \midrule[1.5pt]
      \makecell{ISAC\\with $s_{\sf R}$ and SIC} & $\delta_{\sf SIC}=0$ & $\delta_{\sf SIC}=0,~\mathbf{p}_{\sf c}=\mathbf{0}$\\
      \midrule
      \makecell{ISAC\\with $s_{\sf R}$ and no SIC} & $\delta_{\sf SIC}=1$ & $\delta_{\sf SIC}=1,~\mathbf{p}_{\sf c}=\mathbf{0}$\\
      \midrule
      \makecell{ISAC\\without $s_{\sf R}$} & $\mathbf{p}_{\sf R}=\mathbf{0}$ & $\mathbf{p}_{\sf R}=\mathbf{p}_{\sf c}=\mathbf{0}$\\
      \midrule
      \makecell{Communication\\-only} & \makecell{$\mathbf{p}_{\sf R}=\mathbf{0}$,\\w/o CRB constraints} & \makecell{$\mathbf{p}_{\sf R}=\mathbf{p}_{\sf c}=\mathbf{0}$,\\w/o CRB constraints}\\
      \bottomrule[1.5pt]
      \end{tabular}
      \renewcommand{\arraystretch}{1}
\end{table}
\subsection{Communication Performance}
In Fig.  \ref{Communication_Performance}, we compare the minimum rate performance under eight possible transmission modes: 1) RSMA-based communication-only, 2) RSMA-based ISAC with $s_{\sf R}$ and SIC, 3) RSMA-based ISAC with $s_{\sf R}$ and no SIC, 4) RSMA-based ISAC without $s_{\sf R}$, and 5-8) the same for SDMA. The performance of each mode can be obtained by setting variables, as shown in Table II. For SDMA-based ISAC, using radar sequences and enabling SIC improves performance by a minimum of around $\num{2}$\% (at $P_t = \num{30}$ dBW) and a maximum of $\num{90}$\% (at $P_t = \num{20}$ dBW). However, the performance of each mode of the RSMA-based ISAC remains the same regardless of the use of radar sequence and corresponding SIC. The rate-splitting gain (relative performance gain of RSMA-based ISAC compared to the SDMA-based ISAC with $s_{\sf R}$ and SIC) ranges from a minimum of $\num{23}$\% (at $P_t = 30$ dBW) to a maximum of $\num{76}$\% (at $P_t = 20$ dBW). This demonstrates that RSMA's capability to effectively manage inter-/intra-function interference enables superior communications performance even without a dedicated radar sequence.
\begin{figure*}[!t]
\centering
 \includegraphics[width=1\linewidth]{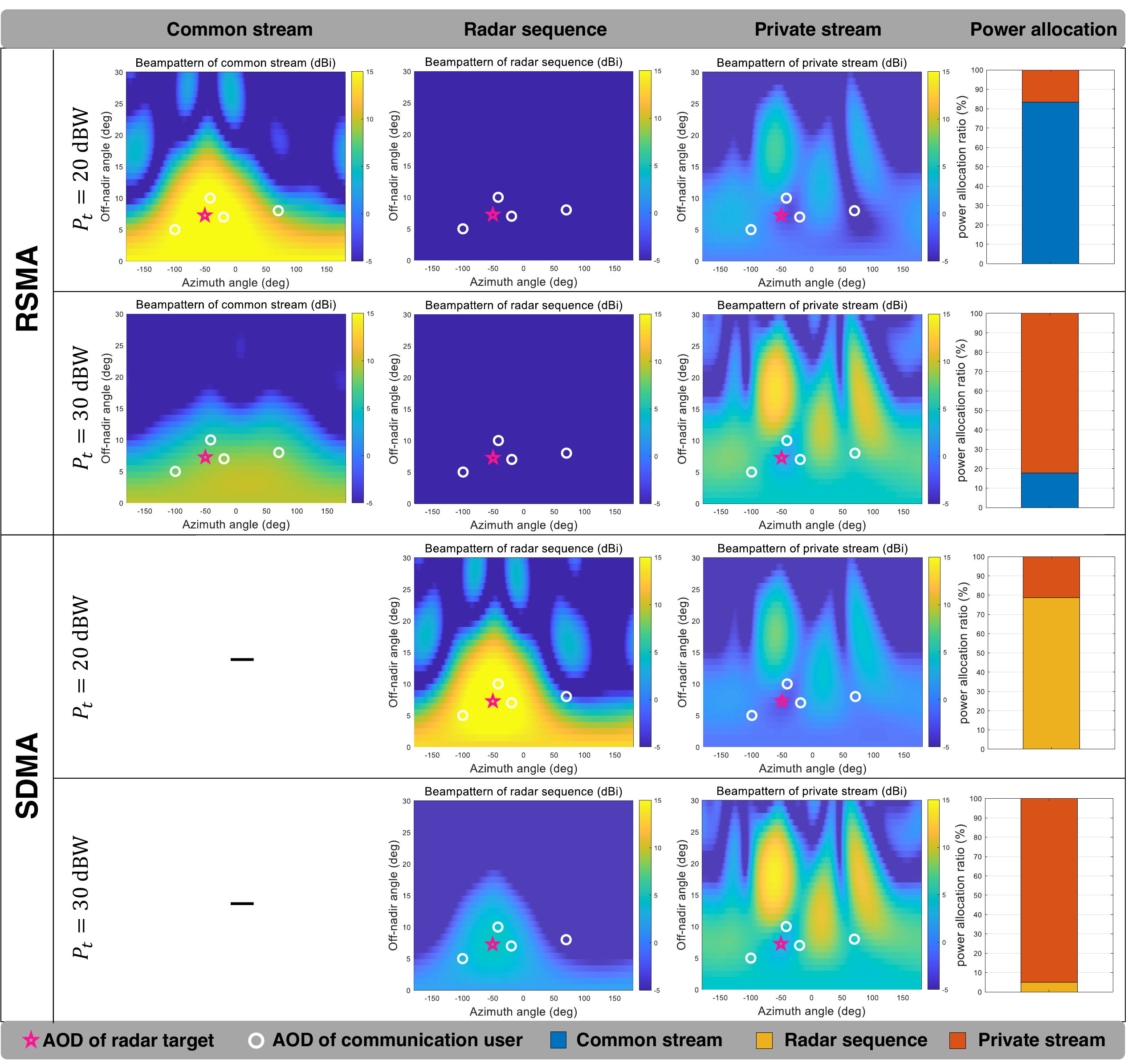}
 		\caption{Transmit beampatterns and power allocation ratios for the radar sequence, common stream, and private stream in RSMA and SDMA when radar sequence and corresponding SIC are enabled with the CRB constraints $\gamma_\theta^{\sf th}=\gamma_\phi^{\sf th}=8\times10^{-7}~{\rm rad}$.}
   \label{beampattern}
    	\vspace{-6mm}
\end{figure*}

Fig. 6 demonstrates that our proposed optimization framework enables the common stream to flexibly and adaptively perform the following three key roles under the limited power budget: i) beamforming towards the radar target, ii) inter-function interference management, and iii) intra-function interference management. To effectively visualize the optimized behavior of the common stream, we present the transmit beampatterns (on the left side) and the power allocation ratios for the common stream, private streams, and the radar sequence (on the right side). For both RSMA- and SDMA-based ISAC, the radar sequence and the corresponding SIC are enabled by setting variables, as shown in Table II. Each beampattern is calculated as follows:
\begin{align}
P_{\sf radar}(\theta,\phi)=\frac{\left|\mathbf{p}_{\sf R}^{\sf H}\mathbf{a}\left(\theta,\phi\right)\right|^2}{\Vert\mathbf{P}\Vert_{\sf F}^2},
\nonumber
\end{align}
\begin{align}
P_{\sf common}(\theta,\phi)=\frac{\left|\mathbf{p}_{\sf c}^{\sf H}\mathbf{a}\left(\theta,\phi\right)\right|^2}{\Vert\mathbf{P}\Vert_{\sf F}^2},
\nonumber
\end{align}
\begin{align}
P_{\sf private}(\theta,\phi)=\frac{\sum_{k\in\mathcal{K}}\left|\mathbf{p}_{k}^{\sf H}\mathbf{a}\left(\theta,\phi\right)\right|^2}{\Vert\mathbf{P}\Vert_{\sf F}^2}.
\nonumber
\end{align}
With the power budget of $\num{20}$ dBW, the majority of the power is allocated to either the common stream (in RSMA-based ISAC) or the radar sequence (in SDMA-based ISAC) to focus a beam to the radar target. This is because the CRB constraints ensure a certain amount of power towards the radar target. To this end, most of the available power budget should be allocated to meet CRB constraints when $P_t = \num{20}$ dBW or lower power budget. Therefore, in RSMA-based ISAC, the common stream mainly plays its first role (beamforming to the target) and naturally undertakes its second role (inter-function interference management) as it will be removed from the users via SIC. By doing so, the common stream takes the place of the radar sequence, resulting in no power allocation to the radar sequence in the RSMA-based ISAC. 

As shown in Fig. 6, when the power budget increases to 30 dBW, the radar sequence in SDMA-based ISAC forms a beam towards the radar target only, while the common stream in RSMA-based ISAC forms a beam towards not only the radar target but also communication users.
To be more specific, when $P_t=\num{30}$ dBW, the power budget is more than enough to satisfy the CRB constraints. 
As such, in SDMA-based ISAC, a relatively small amount of power (about 4\%) compared to that of the case with $P_t=\num{20}$ dBW is allocated to the radar sequence as shown on the right side in Fig. 6. 
In RSMA-based ISAC, on the other hand, much more power is allocated to the common stream (about 18\%). This shows that the common stream primarily serves its third role of intra-function interference management when $P_t = \num{30}$ dBW. In summary, it is confirmed that our proposed framework can achieve superior interference management by adaptively adjusting the power allocation of the common stream due to the flexible nature of our proposed optimization algorithm. By doing so, even without the radar sequence, efficient joint operations can be achieved at LEO satellite systems with limited radio resources. In addition, our observation that there is no need to use an extra radar sequence in RSMA-based ISAC is consistent with the lesson in prior RSMA-based ISAC study \cite{xu2021rate} with an emphasis on sum-rate maximization.






\begin{figure}[!t]
\centering 	\includegraphics[width=0.9\linewidth]{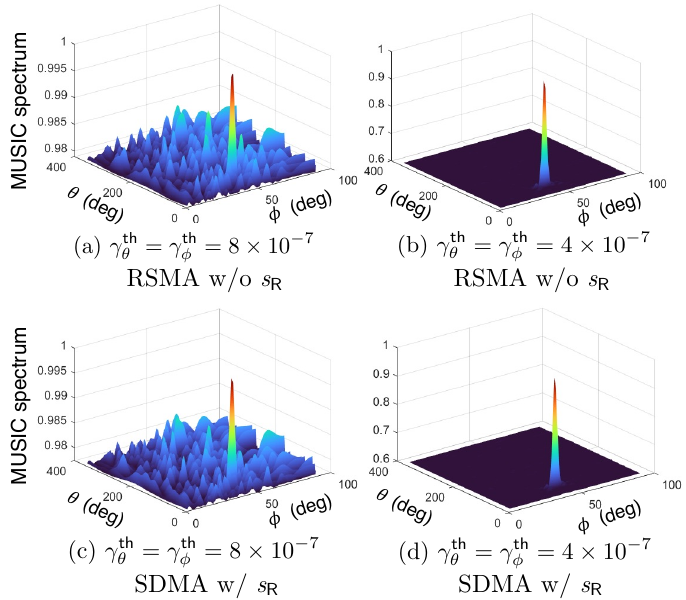}
 		\caption{Normalized MUSIC spectrums for AOA estimation with the total power budget $P_t=23~{\rm dBW}$, (a) and (b): RSMA-based ISAC without $s_{\sf R}$, (c) and (d): SDMA-based ISAC with $s_{\sf R}$.}
    	\label{MUSIC_Spectrum}\vspace{-5mm}
\end{figure}
\begin{figure}[!t]
\centering
 	\includegraphics[width=0.9\linewidth]{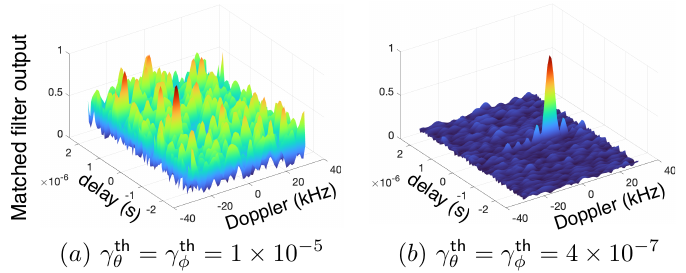}
 		\caption{Matched filter output of the joint delay-Doppler-AOD estimation algorithm with the total power budget $P_t=23~{\rm dBW}$, ${\rm (a)}$: when the previous AOA estimation failed, ${\rm (b)}$: when the previous AOA estimate was accurate.}
\label{Joint_estimation}\vspace{-6mm}
\end{figure}
\begin{remark}
{\rm \textbf{(Comparison of power allocation between the proposed framework and conventional RSMA-based communication-only)}:} In a general RSMA-based communication-only scenario, the power allocation ratio to the common stream tends to increase as the power budget increases since the inter-user interference becomes stronger than the noise \cite{9831440}. 
In our proposed scheme, the power allocation to the common stream rather decreases as the power budget increases, owing to the inclusion of radar functions that require sufficient power towards the radar target for accurate detection.
Specifically, when the power budget is barely sufficient to meet the CRB constraints, most of the power is allocated to the common stream to minimize interference to the user. As the power budget increases, the power allocation to the private stream grows to improve communication performance, which is our new observation.
\end{remark}
\subsection{Radar Performance}
Fig. \ref{MUSIC_Spectrum} shows the MUSIC spectrum for estimating the AOA of the echo signal. By setting more stringent CRB thresholds, the sub-peaks in other locations are reduced, indicating improved radar performance. Furthermore, there is no significant numerical difference in the MUSIC spectrum between RSMA-based ISAC without radar sequence and SDMA-based ISAC with radar sequence. This reveals that \tcb{accurate} radar performance is achieved in  RSMA-based ISAC even without the radar sequence required in SDMA-based ISAC, as the common stream effectively substitutes for the radar sequence.

Fig. \ref{Joint_estimation} illustrates the normalized matched filter output of the proposed joint delay-Doppler-AOD estimation algorithm. Since our algorithm leverages the previously estimated AOA, the delay-Doppler-AOD estimation fails when the AOA is poorly estimated. However, it shows a sharp peak when the AOA is accurately estimated in the previous stage, resulting in accurate delay-Doppler-AOD estimation. Note that the estimation of AOD is automatically successful if the delay is accurately estimated.
\vspace{-2mm}
\section{Conclusion}\label{Sec6}
In this paper, we have explored an RSMA-based bistatic LEO-ISAC. We have formulated the precoder design problem to maximize the minimum rate among all users while satisfying the CRB constraint under the lack of iCSI. We have developed an efficient optimization algorithm by employing the SDR, SCA, and SROCR methods. Numerical results have validated the significantly lower echo path loss and superior interference management capability of our proposed frameworks. By doing so, we have shown that our proposed framework can address the major challenges of LEO-ISAC. We anticipate that our work will play a significant role in developing LEO satellite systems capable of simultaneously performing communication and radar.
\vspace{-3mm}
\bibliographystyle{IEEEtran}
\bibliography{Reff_abb}

\begin{thebibliography}{10}
\providecommand{\url}[1]{#1}
\csname url@samestyle\endcsname
\providecommand{\newblock}{\relax}
\providecommand{\bibinfo}[2]{#2}
\providecommand{\BIBentrySTDinterwordspacing}{\spaceskip=0pt\relax}
\providecommand{\BIBentryALTinterwordstretchfactor}{4}
\providecommand{\BIBentryALTinterwordspacing}{\spaceskip=\fontdimen2\font plus
\BIBentryALTinterwordstretchfactor\fontdimen3\font minus \fontdimen4\font\relax}
\providecommand{\BIBforeignlanguage}[2]{{%
\expandafter\ifx\csname l@#1\endcsname\relax
\typeout{** WARNING: IEEEtran.bst: No hyphenation pattern has been}%
\typeout{** loaded for the language `#1'. Using the pattern for}%
\typeout{** the default language instead.}%
\else
\language=\csname l@#1\endcsname
\fi
#2}}
\providecommand{\BIBdecl}{\relax}
\BIBdecl

\bibitem{MyICCPaper}
J.~Park \emph{et~al.}, ``{RSMA}-based bistatic {ISAC} framework for {LEO} satellite systems,'' in \emph{Proc. IEEE Int. Conf. Commun. Workshops (ICC Workshops)}, 2024, pp. 1--6.

\bibitem{kodheli2020satellite}
O.~Kodheli \emph{et~al.}, ``Satellite communications in the new space era: A survey and future challenges,'' \emph{IEEE Commun. Surv. Tutor.}, vol.~23, no.~1, pp. 70--109, 2021.

\bibitem{xu2021rate}
C.~Xu \emph{et~al.}, ``Rate-splitting multiple access for multi-antenna joint radar and communications,'' \emph{IEEE J. Sel. Top. Signal Process.}, vol.~15, no.~6, pp. 1332--1347, 2021.

\bibitem{liu2022integrated}
F.~Liu \emph{et~al.}, ``Integrated sensing and communications: Toward dual-functional wireless networks for {6G} and beyond,'' \emph{IEEE J. Sel. Areas Commun.}, vol.~40, no.~6, pp. 1728--1767, 2022.

\bibitem{liu2020joint}
F.~Liu \emph{et~al.}, ``Joint radar and communication design: Applications, state-of-the-art, and the road ahead,'' \emph{IEEE Trans. Commun.}, vol.~68, no.~6, pp. 3834--3862, 2020.

\bibitem{terrestrial_ISAC_4}
F.~Liu \emph{et~al.}, ``Cram{\'e}r-{Rao} bound optimization for joint radar-communication beamforming,'' \emph{IEEE Trans. Signal Process.}, vol.~70, pp. 240--253, 2021.

\bibitem{gao2023cooperative}
P.~Gao, L.~Lian, and J.~Yu, ``Cooperative {ISAC} with direct localization and rate-splitting multiple access communication: A {Pareto} optimization framework,'' \emph{IEEE J. Sel. Areas Commun.}, vol.~41, no.~5, pp. 1496--1515, 2023.

\bibitem{chen2022generalized}
L.~Chen \emph{et~al.}, ``Generalized transceiver beamforming for {DFRC} with {MIMO} radar and {MU-MIMO} communication,'' \emph{IEEE J. Sel. Areas Commun.}, vol.~40, no.~6, pp. 1795--1808, 2022.

\bibitem{pucci2022performance}
L.~Pucci \emph{et~al.}, ``Performance analysis of a bistatic joint sensing and communication system,'' in \emph{Proc. IEEE Int. Conf. Commun. Workshops (ICC Workshops)}, 2022, pp. 73--78.

\bibitem{kanhere2021target}
O.~Kanhere \emph{et~al.}, ``Target localization using bistatic and multistatic radar with {5G NR} waveform,'' in \emph{Proc. IEEE Veh. Technol. Conf. (VTC2021-Spring)}, 2021, pp. 1--7.

\bibitem{you2022beam}
L.~You \emph{et~al.}, ``Beam squint-aware integrated sensing and communications for hybrid massive {MIMO} {LEO} satellite systems,'' \emph{IEEE J. Sel. Areas Commun.}, vol.~40, no.~10, pp. 2994--3009, 2022.

\bibitem{liu2024max}
Z.~Liu \emph{et~al.}, ``Max-min fair energy-efficient beam design for quantized {ISAC} {LEO} satellite systems: A rate-splitting approach,'' arXiv preprint arXiv:2402.09253, 2024.

\bibitem{yin2022rate}
L.~Yin and B.~Clerckx, ``Rate-splitting multiple access for dual-functional radar-communication satellite systems,'' in \emph{Proc. IEEE Wireless Commun. Netw. Conf. (WCNC)}, 2022, pp. 1--6.

\bibitem{lee2023coordinated}
J.~Lee \emph{et~al.}, ``Coordinated rate-splitting multiple access for integrated satellite-terrestrial networks with super-common message,'' \emph{IEEE Trans. Veh. Technol.}, vol.~73, no.~2, pp. 2989--2994, 2023.

\bibitem{you2020massive}
L.~You \emph{et~al.}, ``Massive {MIMO} transmission for {LEO} satellite communications,'' \emph{IEEE J. Sel. Areas Commun.}, vol.~38, no.~8, pp. 1851--1865, 2020.

\bibitem{yin2020rate}
L.~Yin and B.~Clerckx, ``Rate-splitting multiple access for multigroup multicast and multibeam satellite systems,'' \emph{IEEE Trans. Commun.}, vol.~69, no.~2, pp. 976--990, 2020.

\bibitem{9424454}
L.~Chen \emph{et~al.}, ``Joint radar-communication transmission: A generalized {P}areto optimization framework,'' \emph{IEEE Trans. Signal Process.}, vol.~69, pp. 2752--2765, 2021.

\bibitem{park2023rate}
J.~Park \emph{et~al.}, ``Rate-splitting multiple access for {6G} networks: Ten promising scenarios and applications,'' \emph{IEEE Netw.}, vol.~38, no.~3, pp. 128--136, 2024.

\bibitem{9831440}
Y.~Mao \emph{et~al.}, ``Rate-splitting multiple access: Fundamentals, survey, and future research trends,'' \emph{IEEE Commun. Surv. Tutor.}, vol.~24, no.~4, pp. 2073--2126, 2022.

\bibitem{chen2023rate}
K.~Chen \emph{et~al.}, ``Rate-splitting multiple access for simultaneous multi-user communication and multi-target sensing,'' \emph{IEEE Trans. on Veh. Technol. (early access)}, 2024.

\bibitem{dizdar2022energy}
O.~Dizdar \emph{et~al.}, ``Energy efficient dual-functional radar-communication: {R}ate-splitting multiple access, low-resolution {DAC}s, and {RF} chain selection,'' \emph{IEEE Open J. Commun. Soc.}, vol.~3, pp. 986--1006, 2022.

\bibitem{daniel2017design}
L.~Daniel \emph{et~al.}, ``Design and validation of a passive radar concept for ship detection using communication satellite signals,'' \emph{IEEE Trans. Aerosp. Electron. Syst.}, vol.~53, no.~6, pp. 3115--3134, 2017.

\bibitem{sayin2019passive}
A.~Sayin, M.~Cherniakov, and M.~Antoniou, ``Passive radar using starlink transmissions: A theoretical study,'' in \emph{Proc. Int. Radar Symp. (IRS)}, 2019, pp. 1--7.

\bibitem{blazquez2022passive}
R.~Bl{\'a}zquez-Garc{\'\i}a \emph{et~al.}, ``Passive radar architecture based on broadband {LEO} communication satellite constellations,'' in \emph{Proc. IEEE Radar Conf. (RadarConf22)}, 2022, pp. 1--6.

\bibitem{leyva2022two}
L.~Leyva \emph{et~al.}, ``Two-stage estimation algorithm based on interleaved {OFDM} for a cooperative bistatic {ISAC} scenario,'' in \emph{Proc. IEEE Veh. Technol. Conf. (VTC2022-Spring)}, 2022, pp. 1--6.

\bibitem{yan2007multitarget}
H.~Yan, J.~Li, and G.~Liao, ``Multitarget identification and localization using bistatic {MIMO} radar systems,'' \emph{EURASIP J. Adv. Signal Process.}, vol. 2008, pp. 1--8, 2007.

\bibitem{kell1965derivation}
R.~E. Kell, ``On the derivation of bistatic {RCS} from monostatic measurements,'' \emph{Proc. IEEE}, vol.~53, no.~8, pp. 983--988, 1965.

\bibitem{roper2022beamspace}
M.~R{\"o}per \emph{et~al.}, ``Beamspace {MIMO} for satellite swarms,'' in \emph{Proc. IEEE Wireless Commun. Netw. Conf. (WCNC)}, 2022, pp. 1307--1312.

\bibitem{sun2023trade}
J.~Sun \emph{et~al.}, ``Trade-off between positioning and communication for millimeter wave systems with {Z}iv-{Z}akai bound,'' \emph{IEEE Trans. Commun.}, vol.~71, no.~6, pp. 3752--3762, 2023.

\bibitem{kay1993fundamentals}
S.~M. Kay, \emph{Fundamentals of statistical signal processing: Estimation theory}.\hskip 1em plus 0.5em minus 0.4em\relax Prentice-Hall, Inc., 1993.

\bibitem{luo2010semidefinite}
Z.-Q. Luo \emph{et~al.}, ``Semidefinite relaxation of quadratic optimization problems,'' \emph{IEEE Signal Process. Mag.}, vol.~27, no.~3, pp. 20--34, 2010.

\bibitem{cao2017sequential}
P.~Cao, J.~Thompson, and H.~V. Poor, ``A sequential constraint relaxation algorithm for rank-one constrained problems,'' in \emph{Proc. Eur. Signal Process. Conf. (EUSIPCO)}, 2017, pp. 1060--1064.

\bibitem{grant2014cvx}
M.~Grant and S.~Boyd, ``{CVX}: Matlab software for disciplined convex programming, version 2.1,'' 2014.

\bibitem{kulikov2015accurate}
G.~Y. Kulikov and M.~V. Kulikova, ``The accurate continuous-discrete extended {Kalman} filter for radar tracking,'' \emph{IEEE Trans. Signal Process.}, vol.~64, no.~4, pp. 948--958, 2015.

\bibitem{MUSIC}
R.~Schmidt, ``Multiple emitter location and signal parameter estimation,'' \emph{IEEE Trans. Antennas Propag.}, vol.~34, no.~3, pp. 276--280, 1986.

\bibitem{marom2022bistatic}
H.~Marom, Y.~Bar-Shalom, and B.~Milgrom, ``Bistatic radar tracking with significantly improved bistatic range accuracy,'' \emph{IEEE Trans. Aerosp. Electron. Syst.}, vol.~59, no.~1, pp. 52--62, 2022.

\end{thebibliography}

\end{document}